\documentclass[prb,twocolumn]{revtex4}% Physical Review B

\usepackage{graphicx}% Include figure files
\usepackage{dcolumn}% Align table columns on decimal point
\usepackage{bm}% bold math

\begin{document}

\title{Temperature and particle size dependence of equilibrium order parameter of FePt}

\author{R. V. Chepulskii$^{1,2,*}$ and W. H. Butler$^{1}$}

\affiliation{$^{1}$Center for Materials for Information
Technology, University of Alabama, Box 870209, Tuscaloosa, Alabama 35487-0209, USA\\
$^{2}$Department of Solid State Theory, Institute for Metal
Physics, N.A.S.U., Vernadsky 36, Kyiv-142, UA-03680, Ukraine}

\date{Submitted 26 April 2005 to Phys.Rev.B}

\begin{abstract}

First, second and third nearest neighbor pair mixing potentials for
equiatomic FePt alloys were calculated from first principles by the
Connolly-Williams method within the canonical cluster expansion
formalism. It was demonstrated that the Connolly-Williams potentials
(based on completely \emph{ordered} states) and the
Korringa-Kohn-Rostoker coherent potential approximation (KKR-CPA)
potentials (based on completely \emph{disordered} state) can be
brought into very close correspondence to each other simply by
increasing the magnitude of the strain-induced interactions added to
the KKR-CPA potential. Using the mixing potentials obtained in this
manner, the dependency of equilibrium L1$_0$ ordering on temperature
was studied for bulk and for (approximately) spherical nanoparticles
ranging in size from 2.5 to 6nm.  The order parameter was calculated
using Monte Carlo simulation and the analytical ring approximation.
The calculated order-disorder temperature for bulk (1495-1514 K) was
in relatively good agreement (4\% error) with the experimental value
(1572K). For nanoparticles of finite size, the (long range) order
parameter changed continuously from unity to zero with increasing
temperature. Rather than a discontinuity indicative of a phase
transition, we obtained an inflection point in the order as a
function of temperature. This inflection point occurred at a
temperature below the bulk phase transition temperature and
decreased as the particle size decreased. Our calculations predict
that 3.5nm diameter particles in configurational equilibrium at
600$^{\circ}$C (a typical annealing temperature for promoting L1$_0$
ordering) have an L1$_0$ order parameter of approximately 0.84
(compared to a maximum possible value equal to unity). According to
our investigations, the experimental absence of (relatively) high
L1$_0$ order in 3.5nm diameter nanoparticles annealed at
600$^{\circ}$C or below is primarily a problem of \emph{kinetics}
rather than equilibrium.

\end{abstract}

\pacs{75.50.Tt, 61.46.+w, 64.70.Nd, 61.66.Dk}

\maketitle

\section{Introduction}

\label{Intro}

Self-assembled, monodispersed FePt nanoparticles are being
intensively investigated for possible future application as an
ultra-high density magnetic storage medium. In order to be useful as
a storage medium, however, these particles, because of their
extremely small volume, $V$,  must have sufficiently high magnetic
anisotropy, $K_u$, to withstand thermal fluctuations of the
direction of magnetization. This requires values of the thermal
stability factor, $(K_u V)/(k_\texttt{B} T)$, of approximately 50.
The particles are usually produced by a "hot soap" process that
yields a disordered f.c.c. solid solution alloy (e.g. Ref.
\onlinecite{Sun}). Such particles are not useful for information
storage in the as-made state because they are paramagnetic at room
temperature due to their low magnetic anisotropy.

Typically, the particles are annealed at a temperature
$T\simeq600^\circ$C in order to induce an ordered L1$_0$
phase\cite{Takahashi03,Sint}.  The layered L1$_0$ phase\cite{L10}
is known from studies of bulk alloys to have an extremely high
magnetic anisotropy ($K_u\cong7\times10^7$ erg/cm$^3$).  This
value of magnetic anisotropy would provide a sufficiently large
thermal stability factor to make 3.5nm diameter particles viable
for information storage.

Unfortunately, it appears to be difficult to a achieve a high
degree of long range atomic order in FePt \emph{nanoparticles}
with $\lesssim$4nm diameter by annealing at
$T\lesssim$600$^{\circ}$C (e.g. Ref.\onlinecite{Takahashi03}). One
can consider two possible reasons for the fact that it has not
been possible to obtain well ordered small particles. First, the
observed order may be low because the particle is \emph{not} in
its equilibrium state due to the slow kinetics at low
temperatures. Alternatively, the \emph{equilibrium} order itself
may be low even at relatively low temperatures because of the
small size of nanoparticles. The latter explanation was suggested
in Ref. \onlinecite{Takahashi03}. There, the order-disorder phase
transition temperature was estimated to decrease with decrease of
particle size. For particle sizes less than 1.5 nm in diameter,
the phase transition temperature was found to be below the typical
annealing temperature ($T\simeq600^{\circ}$C). Therefore, particles of diameter
less than 1.5 nm were predicted to have no long range order in
their equilibrium state at 600$^{\circ}$C. This explanation is in
qualitative agreement with experiment. The difference between the
experimental (4nm) and theoretical (1.5nm) critical size for
disappearance of L1$_0$ order at 600$^{\circ}$C was attributed to
the neglect of nanoparticle surface effects.

From our point of view, however, the results obtained in Ref.
\onlinecite{Takahashi03} require verification because of the
limitations of the theoretical models used in that study. Namely,
the interatomic potentials in alloys usually are much more
complicated and long-ranged than the nearest neighbor
Lennard-Jones model that was used.  In addition, the
order-disorder phase transition temperature was estimated in Ref.
\onlinecite{Takahashi03} by comparing the free energies of
completely ordered and completely disordered states; whereas in
reality, the ordered state approaches (with increasing
temperature) the phase transition point being not completely
ordered. Also, the disordered state would be expected to approach
the phase transition (with decreasing temperature), not with a
completely random atomic distribution but with an atomic
distribution that has substantial short range order.   Moreover,
it is known\cite{NoFinitePhTr} that there is no formal phase
transition in a finite system.

The aim of the present paper is to determine the reason for
the low value observed experimentally for the L$1_0$ order parameter
in small nanoparticles at typical annealing temperatures through the
use of theoretical models that do not have the
above-described limitations.\cite{Short}

Presently, there are two main approaches for calculating effective
atomic interactions (called mixing potentials in this paper) that
determine the configurational behavior of atoms in an alloy. One
is the Connolly-Williams method\cite{CW,Ceder02,CW-name}, which is based
on the first principles calculation of the energies of a number of
completely \emph{ordered} structures. The second approach is based
on the coherent potential approximation (CPA) for the completely
\emph{disordered} state (e.g. Ref. \onlinecite{Ducast}). In the
present paper we utilize and compare both approaches.  The
first-principles calculations that we performed in order to
implement the Connolly-Williams method were carried out in the
generalized gradient approximation to density-functional theory,
using the VASP program package.\cite{VASP} To obtain mixing
potentials from the CPA we used data obtained within the
Korringa-Kohn-Rostoker coherent potential approximation
(KKR-CPA)\cite{Gyorffy-83,Johnson-90,Staunton-94,Pinski-98} in
Ref. \onlinecite{Staunton03}.

To study the temperature, concentration and size dependencies of
equilibrium long-range order in FePt bulk and nanoparticles we used
Monte Carlo simulation (utilizing the Metropolis
algorithm\cite{Metropolis}) and the analytical ring
approximation\cite{Ring}.

\section{Lattice gas model and cluster expansion}

\label{LG}

We consider an Fe-Pt alloy in the framework of the two-component A-B
lattice gas model (e.g. A=Pt, B=Fe). In this model\cite{Lee52}, two
types of atoms are distributed over the sites of a rigid crystal
lattice. The atoms are allowed to be situated only at the crystal
lattice sites and each site can be occupied by only one atom. The
atoms interact through the lattice potentials and can exchange their
positions according to Gibbs statistics. The lattice gas model is
the most commonly used model for describing substitutional ordering
in alloys (e.g. Refs. \onlinecite{Books,Ducast})

The configurational state of the lattice gas considered here can be
defined by the set of configurational variables
$C_{\textbf{R}}^{\alpha}$:
\begin{equation}
\label{C-def}C_{\textbf{R}}^{\alpha}=\left\{
\begin{array}{l}
1
\text{, if the site }{\bf R}\text{ is occupied by an } \alpha \text{-type atom} \\
0
\text{, otherwise}
\end{array},
\right.
\end{equation}
where $\alpha=$A,B and {\bf R} is the site radius-vector. Because
each site contains only one atom, we have
\begin{equation}\label{C}
C_{\textbf{R}}^{\text{A}}+C_{\textbf{R}}^{\text{B}}=1 \Rightarrow
C_{\textbf{R}}^{\text{B}}=1-C_{\textbf{R}}^{\text{A}}.
\end{equation}
Thus any configurational state of the two-component lattice gas
model (i.e. any particular distribution of atoms over the sites)
can be determined by $C_{\textbf{R}}^{\text{A}}$ variables only.

The energy of any state can be expanded as (e.g. Sec. 2 in Ref.
\onlinecite{Sym-I})
\begin{equation}\label{E1}
    E=E_{0}+\sum\limits_{\alpha,{\bf
    R}}E_{\textbf{R}}^{\alpha}C_{\textbf{R}}^{\alpha}
    +\frac{1}{2}\sum\limits_{\alpha_1,\alpha_2}
    \sum\limits_{\textbf{R}_1,\textbf{R}_2}
    E_{\textbf{R}_1,\textbf{R}_2}^{\alpha_1,\alpha_2}
    C_{\textbf{R}_1}^{\alpha_1}C_{\textbf{R}_2}^{\alpha_2},
\end{equation}
where $E_{0}$, $E_{\textbf{R}}^{\alpha}$ and
$E_{\textbf{R}_1,\textbf{R}_2}^{\alpha_1,\alpha_2}$ are the
coefficients of the energy expansion.
$E_{\textbf{R}_1,\textbf{R}_2}^{\alpha_1,\alpha_2}$ can be
considered as pair interactions between two atoms of types
$\alpha_1$ and $\alpha_2$ situated at $\textbf{R}_1$ and
$\textbf{R}_2$ sites, respectively. Expression (\ref{E1}) is usually
called a cluster expansion.\cite{CE} In (\ref{E1}), the terms
proportional to the powers of $C_{\textbf{R}}^{\alpha}$ higher than
second power are not taken into account. Terms corresponding to
non-pair atomic interactions were found to be small in equiatomic
FePt alloy (see discussion in Sec. \ref{CW}) so that we excluded
them from the begining.

By the use of (\ref{C}), one can exclude
$C_{\textbf{R}}^{\text{B}}$ from (\ref{E1}) and get
\begin{equation}\label{E2}
    E=V^{(0)}+\sum\limits_{{\bf R}_{1}}V_{\textbf{R}_{1}}^{(1)}C_{\textbf{R}_{1}}^{\text{A}}
    +\frac{1}{2}\sum\limits_{\textbf{R}_1,\textbf{R}_2}
    V_{\textbf{R}_1,\textbf{R}_2}^{(2)}
    C_{\textbf{R}_1}^{\text{A}}C_{\textbf{R}_2}^{\text{A}},
\end{equation}
where $V^{(i)}$ ($i=0,1,2$) we call mixing potentials
(other names are "interchange energies" and "effective cluster
interactions"):
\begin{equation}\label{V1}
V_{\textbf{R}_{1}}^{(1)}=E_{\textbf{R}_{1}}^{\text{A}}-E_{\textbf{R}_{1}}^{\text{B}}+
\sum\limits_{{\bf
R}_{2}}[E_{\textbf{R}_1,\textbf{R}_2}^{\text{A},\text{B}}-E_{\textbf{R}_1,\textbf{R}_2}^{\text{B},\text{B}}],
\end{equation}

\begin{equation}\label{V2}
V_{\textbf{R}_1,\textbf{R}_2}^{(2)}=
E_{\textbf{R}_1,\textbf{R}_2}^{\text{A},\text{A}}
-2E_{\textbf{R}_1,\textbf{R}_2}^{\text{A},\text{B}}
+E_{\textbf{R}_1,\textbf{R}_2}^{\text{B},\text{B}},
\end{equation}
$V^{(0)}$ is the energy of lattice gas with B atoms only. The
knowledge of the mixing potentials is important because they
determine the energetics of the lattice gas. A number of methods
have been elaborated for calculation of the mixing potentials.

\section{Connolly-Williams method}

\label{CW}

Within the Connolly-Williams method\cite{CW,Ceder02,CW-name}, the
mixing potentials are assumed to have the symmetry of the disordered
state, i.e. to be configurationally independent. Thus,
$V_{\textbf{R}}^{(1)}$ is independent of $\textbf{R}$ and the pair
mixing potential $V_{\textbf{R}_1,\textbf{R}_2}^{(2)}$ depends only
on the difference $\textbf{R}_1-\textbf{R}_2$:
\begin{equation}\label{Sym}
V_{\textbf{R}}^{(1)}=V^{(1)},\quad
V_{\textbf{R}_1,\textbf{R}_2}^{(2)}=V_{\textbf{R}_1-\textbf{R}_2}^{(2)}.
\end{equation}
The pair mixing potential $V_{\textbf{R}}^{(2)}$ can be represented
by its values $V_{s}^{(2)}$ for different coordination shells:
\begin{equation}\label{Shells}
V_{s}^{(2)} \equiv V_{\textbf{R}_s}^{(2)},
\end{equation}
where $\textbf{R}_s$ is the radius-vector connecting any given site
with another site that belongs to the $s$-th coordination shell with
respect to that given site. Generally, we define a coordination
shell for a given site to be all of the sites that interact with a
given site with the same value of the pair mixing potential.

Within the Connolly-Williams method, the energies $E_i$
($i=1,2,\ldots,N_\varepsilon$) of some $N_\varepsilon$
structures (i.e. long-range ordered periodic atomic distributions sometimes called
superstructures; for example L1$_0$)
are calculated by first
principles methods. Writing Eq. (\ref{E2}) for each of those
$N_\varepsilon$ structures, one can get the following system
of linear equations
\begin{equation}\label{E3}
\left\{
\begin{array}{l}
\varepsilon_i=V^{(1)}c_i+\sum\limits_{s=1}^{N_s}S_{is}V_s^{(2)} \\
i=1,2,\ldots,N_\varepsilon
\end{array},
\right.
\end{equation}
where
\begin{equation}\label{Ei}
\varepsilon_i=\frac{E_i-V^{(0)}}{N^{\text{A}}_i+N^{\text{B}}_i},
\end{equation}
$N^{\text{A}}_i$ and $N^{\text{B}}_i$ are the total numbers of A and
B atoms in $i$-th structure, respectively, $S_{ij}$ are the
structural coefficients, and $c_i$ is the concentration of A atoms:
\begin{equation}\label{conc}
c_i=N^{\text{A}}_i/(N^{\text{A}}_i+N^{\text{B}}_i).
\end{equation}
The energy $V^{(0)}$ of a pure B crystal is also calculated by first
principles methods. So the problem is to find $N_s+1$ unknown mixing
potentials $V^{(1)}$ and $V_s^{(2)}$ ($s=1,2,\ldots,N_s$) that give
the best fit [through Eq. (\ref{E3})] to the energies
$\varepsilon_i$ ($i=1,2,\ldots,N_\varepsilon$) calculated by the
first principles methods.

Usually, the choice of the structures to be used is random with a
predisposition towards the well-known ones that are more easily
calculated.  It is preferable that the structures be experimentally
observable for the alloy under consideration and that they
correspond to the ground state or at least a low energy state (e.g.
Ref. \onlinecite{CW,Ceder02}). However, since all configurations are
possible for Eqs. (\ref{E2}) and (\ref{E3}) \textit{any} convenient
structure can be used for determination of the mixing potentials.

So, in contrast to the usual procedure, our criterion for the choice
of the structures is only that they be linearly independent. By
\emph{linear independence} we mean that the main determinant of the
system (\ref{E3}) i.e. the determinant of the matrix
$\|\delta_{1,j}(c_i-S_{ij})+S_{ij}\|$ is not zero
($i,j=1,2,\ldots,N_\varepsilon$; $\delta$ is the Kronecker delta).
Among the linearly independent structures we chose those with less
than twenty atoms per unit cell in order to reduce the cost of the
first principles calculations. Note that if all of the structures
have the same composition (see below), there is only one ground
state (if it is not degenerate) and the usual procedure of using
only (or mainly) the ground states within the Connolly-Williams
method is inapplicable. One may hypothesize that the use of
non-ground states should expand the applicability of the resulting
mixing potentials to higher temperatures.

In the present paper we mainly (except Figs.
\ref{Fig:PhDi_ring},\ref{Fig:PP_c}) study the Fe-Pt alloy close to
equiatomic composition $c=0.5$. Accordingly, we chose input
structures to be of the same composition $c_i=0.5$. The reason for
doing so is the following. If we use structures of different
compositions to find the mixing potentials, we imply that the mixing
potentials are exactly the same for all compositions. However, the
electronic structure of an alloy system generally depends strongly
on composition. (For example, at low temperatures the FePt alloy
shows a strong dependence of magnetic properties on composition:
Fe$_3$Pt, FePt and FePt$_3$ are ferromagnetic, ferromagnetic and
antiferromagnetic, respectively.) The mixing potentials are mainly
determined by the electronic structure of the alloy and, therefore,
one may expect a compositional dependence of the mixing potentials
as well. For example, such a dependence of pair mixing potential was
found numerically within KKR-CPA in Ref. \onlinecite{CPA-V(c)}. For
such systems with strong compositional dependence of properties
one may expect faster convergence of the Connolly-Williams method
results with respect to increase of $N_\varepsilon$ and $N_s$ when
the considered structures are of the same composition. Such an
approach corresponds to the canonical cluster expansion
formalism\cite{CCE} and was applied e.g. in Ref. \onlinecite{Zark}.

Note that the Connolly-Williams method is usually applied using
structures of different compositions assuming compositional
independence of mixing potentials (grand canonical cluster expansion
formalism\cite{CCE}). In this approach, the compositional dependence
of alloy properties (e.g. the asymmetry of the phase diagram with
respect to $c=0.5$) manifests itself through the presence of nonpair
mixing potentials.\cite{PhDiAsym} In our approach the composition
dependence of alloy properties arises partly from the composition
dependence of the mixing potentials. This approach is similar to
that used in methods based on the study of the disordered state
(e.g. CPA). Therefore, a direct comparison of Connolly-Williams and
CPA potentials is possible (see below Sec. \ref{CPA}).

The first principles calculations were performed within the
generalized gradient approximation to density-functional theory,
using the VASP program package with mainly default
settings.\cite{VASP} All calculations were spin polarized. The
effect of lattice vibrations was omitted. All structures were
totally relaxed including shape and volume relaxation of the unit
cell as well as the individual displacements of atoms within the
unit cell. The densities of $k$-points using the Monkhorst-Pack mesh
in the corresponding full Brillouin zones were chosen to be similar
for all considered structures and approximately equal to
$\frac{8\times8\times8}{(2\pi/a)^3}$ ($a$ is the f.c.c. lattice
parameter), but the convergence of the results was checked to verify
that it was sufficient in all cases. For the calculation of
$\varepsilon_i$ [see Eq. (\ref{Ei})] for \textit{each} structure we
used the quantity $V^{(0)}$ (energy of pure Fe) calculated
separately using the same parameters (both unit cell and VASP
parameters) as for the corresponding structure.  We believe that
this approach diminished the error of the \emph{difference}
$E_i-V^{(0)}$ (and correspondingly of $\varepsilon_i$) compared to
the error of $E_i$ and $V^{(0)}$ themselves, due to the systematic
cancelation of errors.

The L1$_0$ structure was included among the structures considered.
In this case, after atom position relaxation, we obtained
$3.848$\AA\, and $3.771$\AA\, for the $a$ and $c$ lattice parameters
of the corresponding tetragonal lattice, respectively ($c/a=0.980$).
For comparison the experimental values are $3.847$\AA\, and
$3.715$\AA\, ($c/a=0.966$).\cite{acExper} In addition, our
calculated results showed the L1$_0$ ferromagnetic structure to be
more stable (i.e. has lower energy) than the antiferromagnetic one
in accordance with experiment. We believe that this good
correspondence between theoretical and experimental results confirms
the adequacy of our VASP first principles calculations.

\begin{figure}
\includegraphics{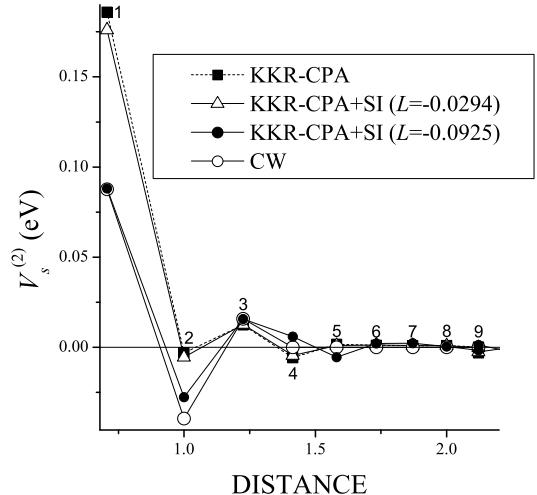}
\caption{The values of pair mixing potential $V_{s}^{(2)}$ for
nine coordination shells ($s=1,2,\ldots,9$) corresponding to the
KKR-CPA method (KKR-CPA), KKR-CPA plus strain-induced interactions
(KKR-CPA+SI) at $L=-0.0294; -0.0925$ (see Sec. \ref{CPA}) as well
as to the Connolly-Williams method (CW) at $N_s=3$, $N_{\varepsilon}=23$
(see Sec. \ref{CW}). The distance is measured in f.c.c. lattice
parameter units.} \label{Fig:V_r}
\end{figure}

\begin{figure}
\includegraphics{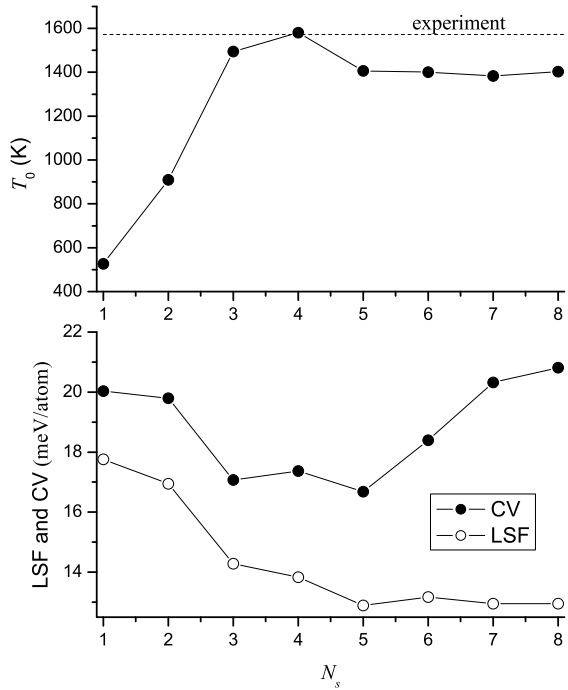}
\caption{The values of cross-validation (CV) and least-squares
fitting (LSF) errors (see Eqs. (\ref{ErrorsCV})-(\ref{ErrorsLSF}))
as well as of the phase transition temperature $T_0$ (within the
ring approximation\cite{Ring}) calculated as a function of the
number of pair mixing potential values $N_s$ taking into account
within the Connolly-Williams method ($N_{\varepsilon}=23$). Dashed
line corresponds to the experimental value\cite{FePt-L10-bulk}.}
\label{Fig:CV}
\end{figure}

\begin{figure}
\includegraphics{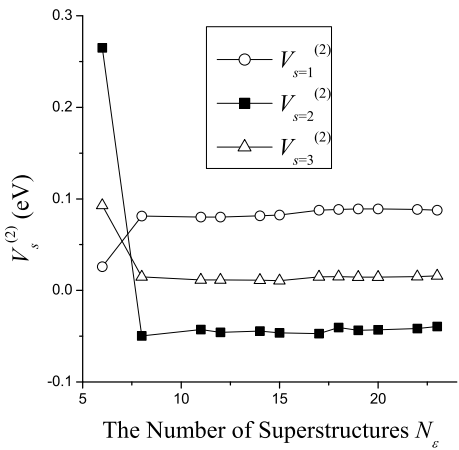}
\caption{The values of pair mixing potential $V_{s}^{(2)}$ for
three coordination shells ($s=1,2,3$) obtained by the
Connolly-Williams method at fixed $N_s=3$ but for different
numbers $N_{\varepsilon}$ of structures (all of equiatomic
composition $c=0.5$) taken into account within the method.}
\label{Fig:Converge}
\end{figure}

\begin{figure}
\includegraphics{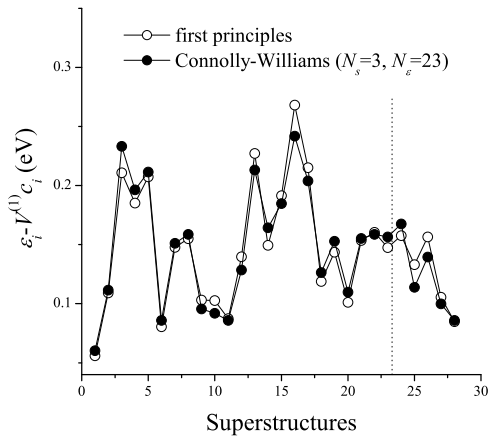}
\caption{The values of $\varepsilon_i-V^{(1)}c_i$ [see Eqs.
(\ref{E3}-\ref{Ei})] for twenty eight equiatomic ($c_i=0.5$)
structures calculated both directly from first principles and by
the use of the mixing potentials obtained by the Connolly-Williams
method (at $N_s=3$, $N_{\varepsilon}=23$). The differences between
those two energies characterize the accuracy of Connolly-Williams
fitting in the case of each structure. The first twenty three
structures (see Appendix \ref{SCW}) are linearly independent and
they (and only they) were used in the Connolly-Williams method.
The last five structures (right side of the vertical dashed line)
correspond to all distinguishable (but not identical to the
previous twenty three ones) equiatomic structures based on
two-cubic unit cells in the f.c.c. crystal lattice (see Appendix
\ref{SCW}). Those five structures were not used in the
Connolly-Williams fitting, but they were considered in order to
assess the predictive power of the obtained Connolly-Williams
potentials.} \label{Fig:CW_E}
\end{figure}

The results of the application of our Connolly-Williams method to
FePt are presented in Figs. \ref{Fig:V_r}-\ref{Fig:CW_E} and Table.
\ref{TAB:V_r}. In Fig. \ref{Fig:CV}, in analogy with Ref. \onlinecite{ZarkPRL},
we present the dependencies of cross-validation\cite{Ceder02} (CV) and least-squares fitting (LSF) errors
\begin{equation}\label{ErrorsCV}
(\text{CV})^2=(23)^{-1}\sum\limits_{i=1}^{23} %
 (\varepsilon_i^{\text{VASP}}-\varepsilon_i^{\text{CW}_i})^2,
\end{equation}
\begin{equation}\label{ErrorsLSF}
(\text{LSF})^2=(23)^{-1}\sum\limits_{i=1}^{23} %
(\varepsilon_i^{\text{VASP}}-\varepsilon_i^{\text{CW}})^2
\end{equation}
as well as of the phase transition temperature on $N_s$. In Eqs.
(\ref{ErrorsCV})-(\ref{ErrorsLSF}): $\varepsilon_i^{\text{VASP}}$
($i=1,2,...,23$) correspond to the values obtained within the VASP
code for 23 linearly independent structures (see Appendix
\ref{SCW}); $\varepsilon_i^{\text{CW}}$ are obtained for those
structures using the mixing potentials obtained by the
Connolly-Williams method at $N_{\varepsilon}=23$;
$\varepsilon_i^{\text{CW}_i}$ are obtained for those structures
using the mixing potentials obtained by the Connolly-Williams
method at $N_{\varepsilon}=22$ when $i$-th structure is excluded
from fitting. Such a CV error characterizes the predictive power
of these Connolly-Williams potentials within the set of 23
structures\cite{Ceder02}.

In Fig. \ref{Fig:Converge},
the convergence of the results obtained by the Connolly-Williams
method at $N_s=3$ is verified with respect to an increase in the number,
$N_{\varepsilon}$, of structures taken into account within the
method. Up to 23 linearly
independent structures (see Appendix \ref{SCW}) were considered.
Note that to achieve
the rapid convergence shown in Fig. \ref{Fig:Converge} it was
important to begin with structures having the highest symmetries.
The use of linearly independent structures guarantees that
the good convergence (starting from $N_{\varepsilon}\simeq8$) is not
caused by a simple similarity of structures.

In Fig. \ref{Fig:CW_E} we compare the values of
$\varepsilon_i-V^{(1)}c_i$ [see Eqs. (\ref{E3}-\ref{Ei})] for each
of the twenty three linearly independent structures (see Appendix
\ref{SCW}).  The figure shows the values calculated both directly
from first principles and by the use of the mixing potentials
obtained by the Connolly-Williams method at $N_s=3$,
$N_{\varepsilon}=23$. The absolute differences between those two
energies characterize the accuracy of Connolly-Williams fitting in
the case of each structure. The average accuracy of 5.72\% (8.63
meV) per one structure was obtained. In order to check the
predictive power of these Connolly-Williams potentials outside the
set of 23 structures, we also calculated the energies of five
additional structures.  This set of five structures includes all
distinguishable (but not identical to the previous twenty three
ones) equiatomic structures based on two-cubic unit cells in the
f.c.c. crystal lattice (see Appendix \ref{SCW}). The average
accuracy of the Connolly-Williams potential for the five
additional structures and for all 28 structures were obtained to
be 7.60\% (10.52 meV) and 6.06\% (8.97 meV) per one structure,
respectively. It is important that if we calculate the
Connolly-Williams potentials using all 28 structures
($N_{\varepsilon}=28$) we obtain almost the same average
accuracies of the energy fitting for the first 23, last five and
all 28 structures: 5.92\% (8.53 meV), 7.79\% (10.36 meV) and
6.25\% (8.85 meV) per one structure, respectively. Thus, one may
conclude that the comparatively lower accuracy of fitting for the
last five structures is not caused by not including them into the
fitting set. Such a high predictive power confirms our choice of
using only linearly independent (see above) structures in the
Connolly-Williams method, because the last five structures in Fig.
\ref{Fig:CW_E} are linearly dependent (but not identical) to the
first twenty three structures. One may suppose that the use of
additional linearly dependent structures in the Connolly-Williams
method would simply impose extra weight factors for some of the
initial structures.

\begin{table}[htbp]
\caption{The values of pair mixing potential $V_{s}^{(2)}$ for
seventeen coordination shells ($s=1,2,\ldots,17$) corresponding to
the KKR-CPA method (KKR-CPA), strain-induced interactions (SI) at
$L=-0.0294$, KKR-CPA method plus
strain-induced interactions at $L=-0.0294$ (KKR-CPA+SI) (see Sec. \ref{CPA}) as well as
calculated by the Connolly-Williams method (CW) at $N_s=3$, $N_{\varepsilon}=23$
(see Sec. \ref{CW}). See also Fig. \ref{Fig:V_r}. The cartesian coordinates of vector $%
{\bf R}$ are given in $a/2$ units, where $a$ is the f.c.c. lattice
parameter. Potential values are in meV units.} \label{TAB:V_r}
\begin{tabular}{ccc|cccc}
Shell & ${\bf R}$ & $\left| {\bf R}\right| /a$ %
                    & KKR-CPA & SI & KKR-CPA+SI & CW \\ \hline
1 & 110 & 0.707 & 185.855 & -9.879 & 175.976 &  87.69 \\
2 & 200 & 1.000 &  -2.997 & -2.484 &  -5.481 & -39.46 \\
3 & 211 & 1.225 &  12.413 &  0.326 &  12.739 &  15.85 \\
4 & 220 & 1.414 &  -5.714 &  1.173 &  -4.541 &   \\
5 & 310 & 1.581 &   1.727 & -0.724 &   1.003 &   \\
6 & 222 & 1.732 &   1.214 &  0.091 &   1.304 &    \\
7 & 321 & 1.871 &   0.907 &  0.133 &   1.040 &   \\
8 & 400 & 2.000 &   0.974 & -0.053 &   0.921 &    \\
9 & 411 & 2.121 &   0.147 & -0.188 &  -0.041 &    \\
  & 330 &       &  -2.997 &  0.438 &  -2.560 &   \\
10 & 420 & 2.236 & -0.050 & -0.209 &  -0.258 &    \\
11 & 332 & 2.345 &  0.015 &  0.030 &   0.045 &    \\
12 & 422 & 2.449 &  0.099 &  0.060 &   0.160 &   \\
13 & 431 & 2.550 & -0.075 &  0.074 &  -0.001 &    \\
   & 510 &       &  0.030 & -0.058 &  -0.028 &    \\
14 & 521 & 2.739 & -0.006 & -0.073 &  -0.079 &    \\
15 & 440 & 2.828 & -0.042 &  0.163 &   0.121 &    \\
16 & 433 & 2.915 &  0.008 &  0.019 &   0.027 &    \\
   & 530 &       & -0.008 & -0.058 &  -0.066 &    \\
17 & 442 & 3.000 & -0.007 &  0.023 &   0.016 &   \\
   & 600 &       &  0.000 & -0.037 &  -0.037 &
\end{tabular}
\end{table}

The good convergence as a function of $N_{\varepsilon}$ shown in Fig. \ref{Fig:Converge},
the high predictive power within and outside the set of 23 structures, the
small CV and LSF errors of the Connolly-Williams fitting shown in  Figs.
\ref{Fig:CV}-\ref{Fig:CW_E}, and the position of the local minimum\cite{CVmin} of CV error as a function of the number of fitting parameters in Fig. \ref{Fig:CV}
suggest that the neglect of nonpair mixing
potentials and of pair mixing potentials outside three coordination
shells (i.e. $N_s=3$)\cite{higher_shells} as well as consideration
of $N_{\varepsilon}=23$ linearly independent structures is
sufficient to obtain adequate results within the Connolly-Williams
method for the case considered here. The small difference between
the experimental bulk order-disorder phase transition temperature
and that obtained using these mixing potentials\cite{Noise} (see Fig.
\ref{Fig:CV} and below Tab.
\ref{TAB:T0} in Sec. \ref{Bulk phase transition}) provides indirect
evidence of adequacy of the potentials for equiatomic Fe-Pt.
In Fig. \ref{Fig:V_r} and Tab. \ref{TAB:V_r} the
pair mixing potential values obtained by the Connolly-Williams method
at $N_s=3$, $N_{\varepsilon}=23$ (four fitting parameters) are shown. Uniform weight factors were used in the fitting.
We obtained also $V^{(1)}=1.394$ eV.

Independent calculations within the KKR-CPA and the
lattice-statics method also suggest that the atomic interaction
tail outside the fourth coordination shell is weak (see below Sec.
\ref{CPA}), despite the "$k\rightarrow0$
non-analyticity"\cite{CW,Khacha,Krivoglaz} of the Fourier
transform of the corresponding pair mixing potential.

In Ref. \onlinecite{Mohri}, the Connolly-Williams method with
structures of different concentrations was applied to FePt.
Consideration was not limited to the ground state structures. One
"hypothetical" phase L$1_1$ was considered. The interchange model
was restricted to two coordination shells of atomic pair
interactions neglecting non-pair interactions. Only six structures
were considered making the number of unknown potential parameters
equal to the number of input parameters (energies of structures).
The convergence of the results obtained by the Connolly-Williams
method with respect to an increase in number of considered
structures was not checked. Consideration of Fig.
\ref{Fig:Converge}, indicates that for such a small number of
structures convergence may not be achieved because of errors in
the energy calculation. For all structures, the relaxation of the
unit cell volume was performed in a way that neglected possible
unit cell shape relaxation (resulting e.g. in its tetragonality)
as well as neglecting the individual displacements of atoms within
the unit cell. Using potentials calculated in this way, the phase
transition temperature for equiatomic FePt bulk alloy was found to
be 2070K, which is much higher than the experimental one
$\sim1573$K.

In Ref. \onlinecite{Seagate}, the ratio
$V_{2}^{(2)}/V_{1}^{(2)}=-0.729$ was derived through the use of
first principles calculations. This is different from the value
$V_{2}^{(2)}/V_{1}^{(2)}=-0.45$ obtained by the Connolly-Williams
method used in the present paper. It is difficult to determine the
reason for this difference because details of the first principles
calculations of Ref. \onlinecite{Seagate} are not reported. It is
possible that the elastic contribution to the mixing potential
neglected in Ref. \onlinecite{Seagate} is responsible for the
difference. In Ref. \onlinecite{Seagate} the value of
$V_{1}^{(2)}=96$ meV (in our designation) was fit by reproducing the
experimental bulk phase transition temperature. This implies a value
of the quantity $V_{2}^{(2)}$ of $=-70$ meV.  For comparison, our
Connolly-Williams potential values (see Tab. \ref{TAB:V_r}) are very
similar for the first coordination shell but almost a factor of two
smaller for the second one. In addition, for the third shell, the
Connolly-Williams potential is nonzero. Of course, the experimental
bulk phase transition temperature is not reproduced perfectly by our
set of Connolly-Williams mixing potentials which was obtained
without fitting to experiment data (see below Tab. \ref{TAB:T0} in
Sec. \ref{Bulk phase transition}).

\section{L1$_0$ parallel antiphase domains}

\label{APB}

For the case of the L1$_0$ structure, there is a possibility of
creating parallel antiphase domains, i.e. the regions where the same
L1$_0$ structures are shifted with respect to one another by the
distance $a/2$ along the $z$-axis perpendicular to the L1$_0$
layers. The antiphase domains are usually created as a result of
nucleation and growth of the L1$_0$ structure in different places as
the initially disordered sample is cooled below the order-disorder
phase transition temperature. The antiphase boundaries (APB) between
the antiphase domains contribute to the total energy making the
existence of antiphase domains favorable or unfavorable depending on
the sign of the antiphase boundary energy.

It is easy to show that such parallel APB energies vanish in the
case of only nearest neighbor interactions. Thus, first principles
calculation of parallel APB energies is important because it allows
one to test the adequacy of the nearest neighbor interaction model
for FePt. In addition, knowledge of the parallel APB energy is
helpful for interpretation of Monte Carlo simulation results (see
below Sec. \ref{MC general}) for the order parameter obtained by
averaging in real space.

In order to estimate the parallel APB energies, we calculated the
values of $E_0$, $E_a$, $E_{2a}$, {\it i.e.} the total energies of
the pure L1$_0$ structure and of two structures composed of
antiphase domains periodically repeated along the $y$ direction with
steps $a$ and $2a$, respectively (corresponding to the structures 1-3 in Appendix \ref{SCW}).
The differences $\delta E_a$ and
$\delta E_{2a}$:
\begin{equation}\label{APB1}
\delta E_a=E_a-E_0,\quad \delta E_{2a}=E_{2a}-E_0
\end{equation}
can be considered as the total energies of a parallel APB in the
corresponding two cases. The results of the calculations are
presented in Tab. \ref{TAB:APB}. Note that, besides the energies
of the parallel APBs themselves, $\delta E_a$ and $\delta E_{2a}$
also contain some contribution from the interaction between
parallel APBs. Such a contribution must decrease with increasing
distance between parallel APBs. The small difference between
$\delta E_a$ and $\delta E_{2a}$ (see Tab. \ref{TAB:APB})
indicates that the interaction between parallel APBs in the  two
cases considered is small and short-ranged. It is somewhat larger
in the relaxed case in accordance with the longer range of elastic
interactions. The positiveness of $\delta E_a$ and $\delta E_{2a}$
indicates that parallel APBs are not energetically favorable. The
nonzero values of $\delta E_a$ and $\delta E_{2a}$ mean that the
nearest neighbor interaction model is not adequate for a
description of FePt alloy in accordance with results obtained in
Sec. \ref{CW}.

\begin{table}
\caption{The energies $\delta E_a$ and $\delta E_{2a}$ of parallel
APBs in the L1$_0$ structure in the cases of $a$ and $2a$ distance
between parallel APBs, respectively. The values shown are per two
atoms in one APB (in eV units) and were calculated by first
principles using the VASP program package.\cite{VASP} In the relaxed
case, we minimized the total energy with respect to the variation of
the size and shape of the unit cell as well as the local
displacements of atoms within the unit cell. In the unrelaxed case,
the atoms were assumed to occupy the sites of a rigid f.c.c. lattice
with the lattice parameters assumed to be the same as in the relaxed
case, neglecting all the local displacements of atoms within the
unit cell.}
\begin{tabular}{c|cc}
   & Unrelaxed & Relaxed \\
  \hline
  $\delta E_a$    & 0.223 & 0.213 \\
  $\delta E_{2a}$ & 0.213 & 0.198 \\
  $\delta E_a-\delta E_{2a}$ & $9.80\star 10^{-3}$ & $1.59\star 10^{-2}$ \\
\end{tabular}
\label{TAB:APB}
\end{table}

\section{KKR-CPA and strain-induced interactions}

\label{CPA}

Elsewhere\cite{Staunton-94,Pinski-98}, detailed procedures have been
described for calculating effective atomic interactions (pair mixing
potentials) within the KKR-CPA
methodology\cite{Gyorffy-83,Johnson-90}. KKR-CPA-based effective
atomic interactions are determined from the response of the
electronic structure of the high-temperature, disordered alloy to
small amplitude concentration waves. The results obtained within
KKR-CPA in Ref. \onlinecite{Staunton03} in the case of
Fe$_{0.5}$Pt$_{0.5}$ alloy are presented in Fig. \ref{Fig:V_r} and
Tab. \ref{TAB:V_r}.

From Fig. \ref{Fig:V_r} and Tab. \ref{TAB:V_r} one may conclude
that the Connolly-Williams and KKR-CPA potentials (both
corresponding to equiatomic concentration $c=0.5$) are similar but
not identical. In both, only the three first coordination shells
are substantially different from zero. The signs of the potential
values for the first three coordination shells are the same within
both methods. However, the absolute value of the Connolly-Williams
potential for the first coordination shell is considerably smaller
than that of the KKR-CPA. For the second shell, the situation is
opposite. One may attribute the difference between the
Connolly-Williams and KKR-CPA potentials to the different
structural and magnetic states used for the calculations:
completely ordered ferromagnetic for the former and completely
disordered paramagnetic for the latter.

For verification of the correspondence of both potentials to the
real FePt alloy we calculated the bulk order-disorder phase
transition temperature (within the ring approximation and by the
Monte Carlo simulation) using both of the calculated mixing
potentials. Then we compared the obtained values with the
corresponding experimental one. The results are presented below in
Tab. \ref{TAB:T0}. One can see that Connolly-Williams and KKR-CPA
potentials underestimate and overestimate the experimental phase
transition temperature, respectively, by almost the same difference.

Within the KKR-CPA calculation, the nuclei were constrained to
occupy ideal f.c.c. crystal lattice positions so the effect of
inhomogeneous static atomic displacements is neglected. (The {\em
homogeneous} ones are taken into account by minimizing the alloy's
total energy with respect to the lattice parameter of the
underlying lattice.) The inhomogeneous static atomic displacements
are expected to be small for FePt alloy because of the small
tetragonality obtained experimentally\cite{acExper} and theoretically
(see e.g. above Sec. \ref{CW}). For verification, we
calculated the corresponding strain-induced contribution to the
pair mixing potential independently within the
Khachaturyan\cite{Khacha} semiphenomenological theory based on the
microscopic
Matsubara-Kanzaki-Krivoglaz\cite{Matsubara,Kanzaki,Krivoglaz}
lattice-statics method. This approach allows us to take into
account the anisotropy and discrete (atomic) structure of a
crystal lattice in contrast with the macroscopic continuum theory
of elasticity.

To calculate the strain-induced interactions, we used the expression
derived in Ref. \onlinecite{Kushwaha} for the Fourier transform of
the dynamical matrix. Platinum and iron were chosen as solvent and
solute, respectively.  The pair mixing potential is
independent of this choice since it is symmetrical with respect to
interchange of A and B, see Eq. (\ref{V2}). For the calculations, we
used the following numerical values: the mass of a Pt atom
$M=32.40\times10^{-26}$ kg; the lattice parameter $a_0=3.92 \AA$;
phonon frequencies: longitudinal $\omega_L=2\pi\times5.80$ and
transverse $\omega_T=2\pi\times3.84$ (all in $\text{Trad s}^{-1}$)
\cite{Omegi}; elastic constants of Pt: $C_{11}=3.580$,
$C_{12}=2.536$, $C_{44}=0.774$ (all in $10^{12}$
dyn/cm$^2$)\cite{ElastConsts}. The coefficient $L$ of linear
dependence of alloy lattice parameter $a$ on the Fe concentration
$c$,
\begin{equation}\label{L}
a=a_0(1+L\, c),\qquad L=\left.\frac{1}{a_0} \frac{\partial
a}{\partial c}\right| _{c=0},
\end{equation}
was calculated as $L=-0.0294$\cite{Pearson}. The concentration
interval used for the calculation of $L$ was chosen so that the
Vegard rule is well fulfilled since it is a necessary condition of
applicability of the Matsubara-Kanzaki-Krivoglaz lattice-statics
method\cite{Khacha,Matsubara,Kanzaki,Krivoglaz}.

The corresponding results for the strain-induced interactions are
presented in Fig. \ref{Fig:V_r} and Tab. \ref{TAB:V_r}. From them it
follows that the strain-induced contribution to the pair mixing
potentials is comparatively small as we expected. The addition of
this contribution to the KKR-CPA potential decreases the
corresponding phase transition temperature so that it is very close
to the experimental one (see Tab. \ref{TAB:T0}).

\begin{figure}
\includegraphics{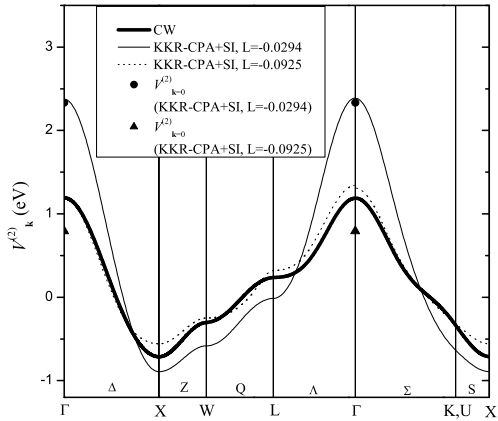}
\caption{The dependencies of the Fourier transform
$V_{\textbf{k}}^{(2)}=\sum\limits_{\textbf{R}}V_{\textbf{R}}^{(2)}\exp(i\textbf{kR})$
of pair mixing potential $V_{\textbf{R}}^{(2)}$ on the wave-vector
along the high-symmetry directions\cite{Bradley and Cracknell 1972}
within the corresponding first Brillouin zone. The three cases
correspond to (1) the pair mixing potentials obtained by the
Connolly-Williams method (CW) at $N_s=3$,
$N_{\varepsilon}=23$ (see Sec. \ref{CW})
(2) pair mixing potentials obtained within the KKR-CPA and (3)
KKR-CPA plus strain-induced interactions (KKR-CPA+SI) at $L=-0.0294;
-0.0925$ (see Sec. \ref{CPA}). The big bold point and triangle
correspond to $V_{\textbf{k}=0}^{(2)}$ in two cases of taking into
account the strain-induced interactions, when the Fourier transform
demonstrates
nonanaliticity\cite{Khacha,Matsubara,Kanzaki,Krivoglaz}.}
\label{Fig:V_k}
\end{figure}

From Fig. \ref{Fig:V_r} and Tab. \ref{TAB:V_r} one can also see that
the addition of the strain-induced contribution to the KKR-CPA pair
mixing potential moves it toward the Connolly-Williams potential
values. To emphasize this effect, we increased the absolute value of
the coefficient $L$ from 0.0294 to 0.0925 (thus artificially
increasing the strain-induced interactions) in order to bring the
KKR-CPA plus strain-induced potential as close to the
Connolly-Williams one as possible. The same tendency is observed in
the Fourier transforms of the corresponding pair mixing potentials
(see Fig. \ref{Fig:V_k}). One may conclude that the mixing potential
calculated for the disordered state (KKR-CPA) can be brought into
close agreement with the mixing potential based on the completely
ordered states (Connolly-Williams method) simply by increasing the
magnitude of the strain-induced interactions (which is proportional
to $L^2$). Perhaps such an increase of the strain-induced
interactions roughly describes the dependence of mixing potential on
the order parameter (configurational excitations of the electronic
structure\cite{Zark,V_PP}).

Note from Tab. \ref{TAB:T0} that the phase transition temperature
corresponding to the KKR-CPA potential plus the artificially
increased strain-induced interactions is much lower than the
experimental one. This may be explained by the roughness of our
approach. Namely, by bringing the KKR-CPA $V_{\textbf{k}}^{(2)}$ to
the Connolly-Williams one in the vicinity of $G$ point (where the
difference between them is largest), we may cause too large an error
in $V_{\textbf{k}}^{(2)}$ in the vicinity of X point, which is
mainly responsible for the phase transition because of minimum of
$V_{\textbf{k}}^{(2)}$ there (see Fig. \ref{Fig:V_k})\cite{magneto}.

From Fig. \ref{Fig:V_r} and Tab. \ref{TAB:V_r} it follows that the
strain-induced potential is of short range despite the
"$k\rightarrow0$ nonanaliticity"\cite{CW,Khacha,Krivoglaz} of its
Fourier transform. One may consider this fact as an additional proof
of correctness of our application of the Connolly-Williams method in
Sec. \ref{CW} using the short range pair mixing potential.

Below, in our Monte Carlo simulation of finite small systems, we
exclusively used mixing potentials obtained by the Connolly-Williams
method. We did so because the most interesting (for the aim of the
present paper) temperature region corresponds to the highly ordered
state. For this reason, we expect that the Connolly-Williams
potentials which are based on the completely ordered state should
work better in that region than the KKR-CPA potentials which are
based on the completely disordered state.

In Ref. \onlinecite{TB} the values of pair mixing potentials for
the FePt bulk alloy were obtained within the tight binding linear
orbital method. For $c=0.5$, the values for the second and fourth
coordination shell mixing potentials are close to the
corresponding KKR-CPA ones from Ref. \onlinecite{Staunton03} but
they are approximately a factor of two smaller for the first and
third shells. Order-disorder transition temperatures of 662K,
816K, and 435K were obtained by us for $c=0.25$, $c=0.50$, and
$c=0.75$, respectively, using these potentials within the ring
approximation.\cite{Ring} These transition temperatures are much
lower than the experimental ones (1087K, 1572K, and 1622K,
respectively)\cite{FePt-L10-bulk}. The corresponding transition
temperatures calculated in Ref. \onlinecite{TB} are not so
different from the experimental ones because the mean-field
approximation, which usually overestimates the transition
temperature\cite{Books}, was used.  The magnitude of the
discrepancy with experiment was underestimated also because an old
value for the transition temperature at $c=0.75$ from Ref.
\onlinecite{Hansen} was used for comparison.

\section{Monte Carlo simulation}

\label{MC general}

For the Monte Carlo simulations we utilized the standard Metropolis
algorithm\cite{Metropolis}. We applied free and periodic boundary
conditions for the cases of spherical nanoparticles\cite{sphere} and
parallelepipeds, respectively.

To obtain a rapid approach to the \emph{equilibrium} state, which is
the subject of this paper, the starting configuration for each
temperature was chosen to be the completely ordered one. We also
used simplified kinetics in our Monte Carlo simulation. Namely, we
allowed \emph{any} two randomly chosen atoms (not only nearest
neighbors) to exchange their positions \emph{without} an additional
diffusion barrier.

The absence of a monotonic change of mean values of all quantities
being calculated in the Monte Carlo simulation was chosen as a
criterion for the achievement of an equilibrium state. To fulfil
such a criterion, it was generally necessary to perform 1000 - 20000
Monte Carlo steps (i.e. the interchange of two atoms chosen at
random) per site. After the equilibrium state was achieved, each
calculated quantity was averaged for subsequent Monte Carlo steps.
When the oscillation amplitudes of such averages (considered as a
function of the number of Monte Carlo steps performed) became less
than 5\% (of the absolute value of the corresponding average) during
the last 10\% of steps (from the total number of steps carried out
at an equilibrium state), the values of the averages calculated at
the last Monte Carlo step were accepted as equilibrium ones.

\begin{figure}
\includegraphics{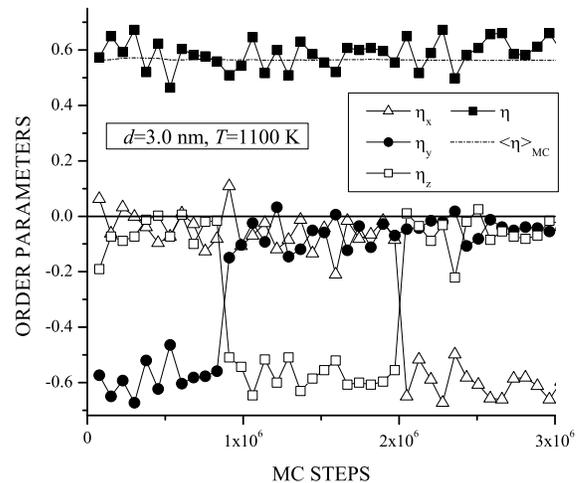}
\caption{The dependence of the FePt L1$_0$ order parameters on Monte
Carlo steps during averaging in the equilibrium state during
simulation of a spherical nanoparticle with diameter $d=3.0$nm (see
Tab. \ref{TAB:Data}) at temperature $T=1100$K.
$\eta_x,\eta_y,\eta_z$ are the "directional" order parameters (see
text), $\eta$ is defined in Eq. (\ref{LRO-def}).} \label{Fig:PPxyz1}
\end{figure}

\begin{figure}
\includegraphics{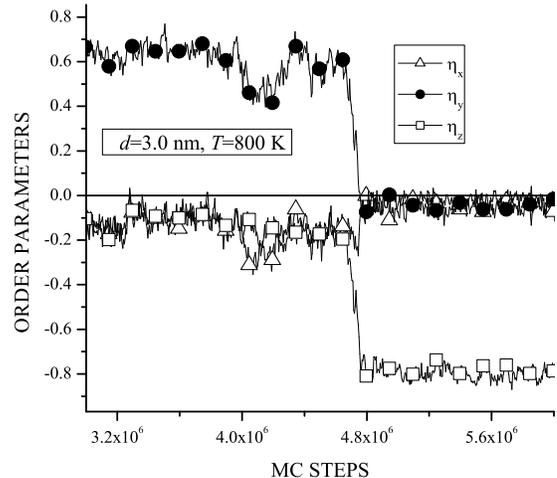}
\caption{The same as in Fig. \ref{Fig:PPxyz1} but for the
"directional" order parameters as the spherical nanoparticle with
diameter $d=3.0$nm (see Tab. \ref{TAB:Data}) approaches its
equilibrium state at temperature $T=800$K.} \label{Fig:PPxyz2}
\end{figure}

We define the equilibrium L1$_0$ order parameter $\overline{\eta}$
as the statistical average of the maximum value among three absolute
values of "directional" order parameters $\eta_x,\eta_y,\eta_z$:
\begin{equation}\label{LRO-def}
\overline{\eta}=\left \langle
\eta\right\rangle_{\texttt{MC}},\quad
\eta=\max\{\left|\eta_x\right|,\left|\eta_y\right|,\left|\eta_z\right|\}
\end{equation}
where $\eta_i\quad(i=x,y,z)$ is defined as the difference between
the Fe atom concentrations at odd and even crystal planes
perpendicular to $i$-th direction,
$\left\langle\ldots\right\rangle_{\texttt{MC}}$ is the statistical
average over the Monte Carlo steps. We chose this definition of
$\eta$ because of the equivalence by symmetry of the $x,y$, and $z$
directions of L1$_0$ order. In addition, one can obtain an
equivalent structure (at $c=0.5$) by changing the sign of $\eta_i$,
which results in the exchange of Fe and Pt atoms producing a
configuration that is equivalent by symmetry to the original one.
During Monte Carlo simulation, we observed fluctuations that cause
transformations between these equivalent states (i.e. fluctuations
in the sign and direction of $\eta$, see Figs.
\ref{Fig:PPxyz1}-\ref{Fig:PPxyz2}). This is in addition to the usual
statistical fluctuations within one such state. The L1$_0$ order
parameter $\overline{\eta}$, defined in Eq. (\ref{LRO-def}) takes
into account the fluctuation induced transformations between the
equivalent states. Note that because of the above-discussed symmetry
equivalence, we obtain $\left \langle
\eta_i\right\rangle_{\texttt{MC}}=0$ for any $i=x,y,z$ at \emph{any}
temperature, when the statistical average is taken over a
sufficiently large number of Monte Carlo steps. Note that in Ref.
\onlinecite{Seagate} an order parameter similar to Eq.
(\ref{LRO-def}) was introduced for the same reasons.

Usually during Monte Carlo simulation we observed one of the
"directional" order parameters to be much larger than other two
(see e.g. Figs. \ref{Fig:PPxyz1}-\ref{Fig:PPxyz2}). This means that
the presence of perpendicular anti-phase domains is not favorable.
Examination of the atomic arrangements within the Monte Carlo
simulation showed no evidence of parallel anti-phase boundaries such
as those discussed in Sec. \ref{APB}. The absence of anti-phase
domains confirms the validity of our definition of the equilibrium
L1$_0$ order parameter in Eq. (\ref{LRO-def}).

Note that even with our simplified kinetics (see above Sec.
\ref{MC general}), we observed a slowing down problem in
approaching the equilibrium ordered state at low temperatures. For
example, in Fig. \ref{Fig:PPxyz2} one can see that, for a long
time, the nanoparticle can be in a metastable state with order
parameter ($\sim$0.6) lower than that ($\sim$0.8) in the
equilibrium state.

\section{Bulk phase transition}

\label{Bulk phase transition}

\begin{figure}
\includegraphics{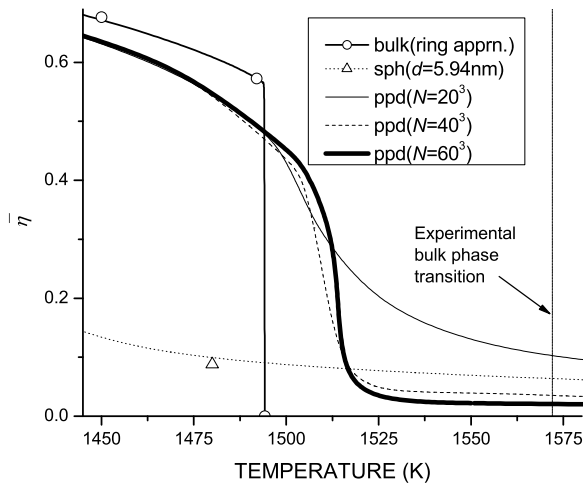}
\caption{The temperature dependence of the FePt equilibrium L1$_0$
order parameter $\overline{\eta}$ obtained within the analytical
ring approximation\cite{Ring} for  bulk ("bulk") as well as by
Monte Carlo simulation for parallelepiped ("ppd") samples
containing $N=20^3,40^3$, and $60^3$ atoms and the spherical
("sph") nanoparticle with $d=5.94$nm diameter (see Tab.
\ref{TAB:Data}) at (or near) equiatomic concentration $c=0.5$.}
\label{Fig:To}
\end{figure}

To obtain an additional verification of the calculated values of
mixing potential, we calculated the order-disorder phase transition
temperature in the \emph{bulk} FePt equiatomic alloy using these
values. To do so, we calculated the temperature dependence of the
FePt equilibrium L1$_0$ order parameter within the analytical ring
approximation\cite{Ring} for bulk as well as by Monte Carlo
simulation for parallelepiped samples containing $N=20^3,40^3$, and
$60^3$ atoms and for spherical nanoparticles with 6.0nm diameter at
(or near) equiatomic composition $c=0.5$. The
results are presented in Fig. \ref{Fig:To}.

From Fig. \ref{Fig:To} one may conclude the following. The ring
approximation (which exactly corresponds to bulk, i.e. to an
infinite sample) clearly shows a phase transition temperature at
which the order parameter $\overline{\eta}$ drops to zero. Strictly
speaking, in all the cases considered here of finite size samples
(sphere and parallelepipeds) there is no phase transition in
accordance with a general theorem\cite{NoFinitePhTr}. The order
parameter $\overline{\eta}$ continuously changes from unity to zero
with increasing temperature and instead of a phase transition we
obtain an inflection point in the $\overline{\eta}(T)$ curve. In the
case of the parallelepiped with $60^3=216000$ atoms, the inflection
point is very similar to the phase transition. Comparing the three
curves for parallelepipeds of different sizes one can imagine that
the inflection point transforms into a phase transition in the limit
of infinite size. The position of the inflection point is often used
to approximate the bulk phase transition point in Monte Carlo
simulations of finite size samples.

According to our calculations, the inflection point corresponding to
the largest parallelepiped simulation can be considered to be a good
estimate for the phase transition within the Monte carlo method. We
plotted the temperature of the inflection point, $T_N$, as a
function of $N^{-1/2}$ and observed the limiting value as
$N^{-1/2}\rightarrow0$.  This extrapolation of $T_N$ to $N\to
\infty$ hardly differs from the temperature of the inflection point
for $N=60^{3}$.

From  Fig. \ref{Fig:To} it follows that Monte Carlo simulation for
parallelepipeds with periodic boundary conditions ("no surface") is much more
appropriate to the bulk behavior than Monte Carlo simulation for the
spherical nanoparticle with free boundary conditions. The number of
particles $N=20^3=8000$ in the smallest parallelepiped considered is
very close to the number $N=8007$  of particles in the spherical
nanoparticle (see below Tab. \ref{TAB:Data}). This large difference
in Monte Carlo simulation between parallelepiped and sphere is
caused by the difference in shape and boundary conditions.

\begin{table}[htbp]
\caption{The values of \emph{bulk} order-disorder phase transition
temperature measured experimentally
(EXPERIMENT)\cite{FePt-L10-bulk} and obtained by the use of mixing
potentials $V_{s}^{(2)}$ corresponding to the KKR-CPA method
(KKR-CPA), KKR-CPA method plus strain-induced interactions at
$L=-0.0294; -0.0925$ (KKR-CPA+SI), Connolly-Williams method (CW),
see Tab. \ref{TAB:V_r}. The theoretical values were calculated
within the ring approximation (ring) and by the Monte Carlo
simulation (MC).} \label{TAB:T0}
\begin{tabular}{c|c}
$V_{s}^{(2)}$ & $T_0$ (K)\\ \hline
EXPERIMENT & 1572 \\
CW         & 1495(ring), 1514(MC)\\
KKR-CPA    & 1610(ring) \\
KKR-CPA+SI ($L=-0.0294$) & 1552(ring)\\
KKR-CPA+SI ($L=-0.0925$) &  997(ring)
\end{tabular}
\end{table}

The temperatures, 1495 K and 1514 K, estimated for the phase
transition within the analytical ring approximation\cite{Ring} and
Monte Carlo simulation, respectively, are in close correspondence to
the experimental\cite{FePt-L10-bulk} 1572 K. This 4\% error
demonstrates the adequacy of the mixing potential values
at $c=0.5$ calculated from first principles (i.e. without a fitting to experimental data).

\begin{figure}
\includegraphics{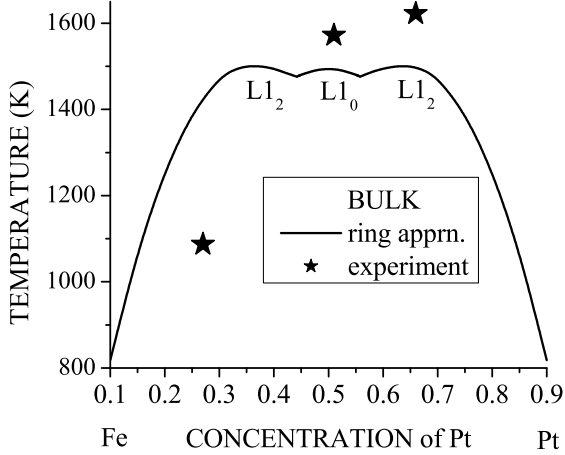}
\caption{Bulk configurational $T-c$ phase diagram obtained within
the analytical ring approximation\cite{Ring} using mixing potentials
calculated by the  Connolly-Williams method at $c=0.5$.
Lines correspond to phase transitions from disorder to L1$_0$ or
L1$_2$ structures with decreasing temperature. Stars correspond to
the phase transitions observed
experimentally.\cite{FePt-L10-bulk}(see also Fig.
\ref{Fig:PhDi_exper})} \label{Fig:PhDi_ring}
\end{figure}

\begin{figure}
\includegraphics{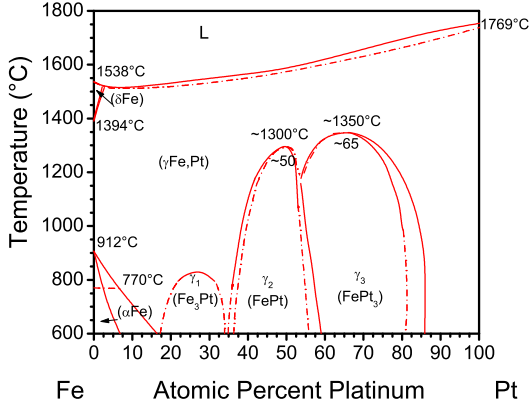}
\caption{Experimental bulk configurational $T-c$ phase
diagram.\cite{FePt-L10-bulk}}
\label{Fig:PhDi_exper}
\end{figure}

In order to test the applicability of these values over the entire
concentration interval, we calculated the order-disorder phase
transition temperature as a function of concentration within the
analytical ring approximation\cite{Ring} using the mixing
potential calculated for $c=0.5$. The corresponding results
together with experimental ones are presented in Fig.
\ref{Fig:PhDi_ring} (see also Fig. \ref{Fig:PhDi_exper}). From
this figure one may conclude that the equiatomic mixing potential
is only valid in the vicinity of $c=0.5$. Outside this vicinity it
is necessary to use different mixing potentials. The asymmetry of
experimental phase diagram with respect to $c=0.5$ implies that
the mixing potentials should be concentration dependent and/or
nonpair.\cite{PhDiAsym} According to our canonical cluster
expansion formalism\cite{CCE}, within the  Connolly-Williams
method one should calculate the mixing potentials at any given
concentration using structures with the \emph{same} (or close)
compositions. Within the KKR-CPA, the concentration dependence of
mixing potentials is obtained naturally.

\section{Nanoparticle simulation}

\label{nanoMC}

The mixing potentials calculated in Sec. \ref{CW} by the
Connolly-Williams method were used for Monte Carlo simulations of
the temperature dependence of the equilibrium L1$_0$ order parameter
of spherical FePt nanoparticles with diameters of 5.94nm, 3.5nm,
3.0nm, and 2.5nm. In all cases the compositions are close to $c=0.5$.
The characteristics of those nanoparticles are listed in Tab.
\ref{TAB:Data}.

\begin{table}
\caption{The characteristics of spherical nanoparticles used in
Monte Carlo simulation: $d$ - diameter; $d/a$ - ratio of diameter
to lattice parameter; $N$, $N_{\texttt{Fe}}$, and
$N_{\texttt{Pt}}$ - the numbers of sites, iron and platinum atoms
in the sample, respectively; $c=N_{\texttt{Fe}}/N_{\texttt{Pt}}$ -
concentration; $\overline{\eta}$ and $D$ - the value and
dispersion (due to the thermodynamic fluctuations) of order
parameter at $600^{\circ}$C, see Fig. \ref{Fig:PP_T_nano};
$c_{\texttt{max}}$ - the value of concentration for which the
concentrational dependence of L1$_0$ equilibrium order parameter
achieves maximum, see Fig. \ref{Fig:PP_c}.}
\begin{tabular}{ccccccccc}
  $d$(nm) & $d/a$ & $N$ & $N_{\texttt{Fe}}$ & $N_{\texttt{Pt}}$ & $c$ & $\overline{\eta}$ & $D$ & $c_{\texttt{max}}$ \\
  \hline
   5.94 & 15.63 & 8007 & 4007 & 4000 & 0.500 & 0.863 & 0.007 & 0.52\\
   3.50 &  9.21 & 1601 &  825 &  776 & 0.515 & 0.839 & 0.021 & 0.53 \\
   3.00 &  7.89 & 1055 &  519 &  536 & 0.492 & 0.746 & 0.033 &  \\
   2.50 &  6.58 &  603 &  299 &  304 & 0.496 & 0.704 & 0.045 & 0.56 \\
\end{tabular}
\label{TAB:Data}
\end{table}

Note that the starting configuration for all cases considered was
chosen to be a completely ordered one with an L1$_0$ order
parameter of unity. The deviations of concentration from
equiatomic $c=0.5$ are due to the finite sizes and spherical shape
of the nanoparticles.

\begin{figure}
\includegraphics{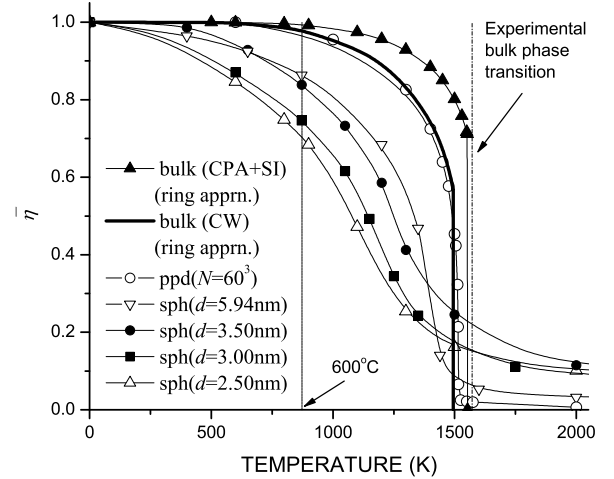}
\caption{The temperature dependence of the FePt equilibrium L1$_0$
order parameter $\overline{\eta}$ obtained within the analytical
ring approximation\cite{Ring} for bulk ("bulk") using KKR-CPA plus
strain-induced interactions at $L=-0.0294$ (CPA+SI) and
Connolly-Williams mixing potential at $N_s=3$,
$N_{\varepsilon}=23$ (CW)
 as well as by Monte
Carlo simulation for the parallelepiped ("ppd") sample containing
$N=60^3$ atoms and for spherical ("sph") nanoparticles with 2.5nm,
3.0nm, 3.5nm, 6.0nm diameters $d$ at equiatomic (or near equiatomic,
see Tab. \ref{TAB:Data}) concentration $c=0.5$ using Connolly-Williams mixing potential.}
\label{Fig:PP_T_nano}
\end{figure}

The results of the simulations are presented in Fig.
\ref{Fig:PP_T_nano}. For spherical nanoparticles, the tendencies in
the behavior of order parameter are the same as in case of
parallelepipeds (see above Sec. \ref{Bulk phase transition}).
Namely, in case of finite size particles, instead of a phase
transition point (see bulk curve) there is an inflection point. The
smaller the particle size the smoother the $\overline{\eta}(T)$
curve, the less order parameter at low temperatures (below and
somewhat above the inflection point) and the higher the order
parameter at high temperatures (well above the inflection point).
The inflection point temperature decreases with decrease of particle
size. Note that in Ref. \onlinecite{Seagate} the same tendencies
were observed but the position of the heat capacity maximum was
interpreted as a phase transition point. Such an interpretation
should be very approximate in case of small finite systems because
there is no formal phase transition in a finite
system\cite{NoFinitePhTr}. The same tendencies as in Fig.
\ref{Fig:PP_T_nano} were observed in Monte Carlo simulation of Cu$_3$Au
nanoparticles in Ref. \onlinecite{Cu3Au}

Our calculations predict, for example, that 3.5nm diameter
particles in configurational equilibrium at 600$^{\circ}$C would
have an order parameter $\overline{\eta}=0.84$ compared to a
maximum possible value of unity (see Tab. \ref{TAB:Data} and Fig.
\ref{Fig:PP_T_nano}). Therefore, annealing at 600$^{\circ}$C will
not yield perfect order for 3.5nm diameter particles.
Approximately 17\% of the atoms will be on the wrong sublattices,
even in equilibrium. The corresponding dispersion of $\eta$ due to
the thermodynamic fluctuations is comparatively small (e.g. 2.5\%
for $d=3.5, T=600^{\circ}$C, see Tab. \ref{TAB:Data}).

The "tail" of the order parameter at high temperatures in Fig.
\ref{Fig:PP_T_nano} is a consequence of thermodynamic fluctuations
of the order parameter. The asymptotic behavior of the dispersion of
such fluctuations can be estimated as $\sim 1/\sqrt{N}$ ($N$ is the
number of atoms)\cite{LandLifsh}. So the value of the high
temperature "tail" of the order parameter increases with decreasing
particle size. Note that we define the order parameter as in Eq.
(\ref{LRO-def}), taking into account the fact that two
configurations with different signs of the order parameter are
physically identical.

\begin{figure}
\includegraphics{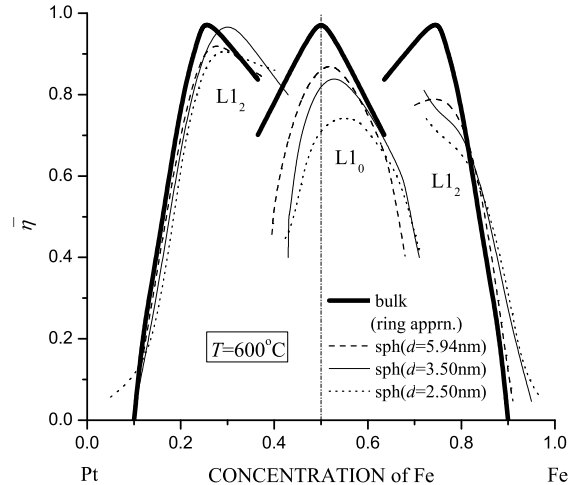}
\caption{The concentration dependence of the FePt equilibrium
L1$_0$ and L1$_2$ order parameters in bulk ("bulk") and spherical
("sph") nanoparticles with 2.5nm, 3.5nm, and 5.94nm diameters $d$
at fixed temperature $T=600^{\circ}$C. The results for bulk and
spherical nanoparticles were obtained within the analytical ring
approximation\cite{Ring} and by Monte Carlo simulation,
respectively, using mixing potentials calculated at $c=0.5$ within
the  Connolly-Williams method.} \label{Fig:PP_c}
\end{figure}

Experimentally, nanoparticles are created with some dispersion in
the concentration. Therefore it is important to know the
concentration dependence of the order parameter. We studied this
dependency for the cases of bulk and spherical nanoparticles of
different diameters at fixed temperature $T=600^{\circ}$C. The
corresponding results are presented in Fig. \ref{Fig:PP_c}. At each
concentration we present the \emph{equilibrium} order parameter
value so that near concentrations $c=0.5$ and $c=0.25; 0.75$ the
curves correspond to L1$_0$ and L1$_2$ order parameters,
respectively.

From Fig. \ref{Fig:PP_c} it follows that for small FePt
nanoparticles we observe an asymmetry in the concentration
dependence of the order parameter with respect to equiatomic
$c=0.5$. This asymmetry is a consequence of the finite size of
nanoparticles. It should be emphasized that for the calculation of
order parameters in Fig. \ref{Fig:PP_c} at each concentration we
used the \emph{same} pair mixing potentials as at $c=0.5$ that
results in a symmetric order parameter for case of bulk. The
concentration dependence of mixing potentials and/or appearance of
non-pair mixing potentials at $c\neq0.5$ can make the bulk curve
asymmetrical\cite{PhDiAsym} and make an additional contribution to
the asymmetry for finite particles observed in Fig.
\ref{Fig:PP_c}. One should consider the results in Fig.
\ref{Fig:PP_c} to be more approximate the further they are from
$c=0.5$.

\section{Surface effects}

\label{Surface effects}

In our study, we have used mixing potentials obtained for
\emph{infinite} bulk alloys and \emph{free} boundary conditions to
simulate the equilibrium configuration of \emph{finite} size particles. The
presence of the surface will change the atomic potentials in the
near-surface region compared to bulk potentials. This change may
result in surface segregation, leading, for example, to the tendency
of Fe or Pt atoms to be preferably situated at the surface. In
general, surface segregation should decrease the total L1$_0$ order
parameter in nanoparticles, because the L1$_0$ order will be reduced
at the surface. The surface effect on the order will depend on the
ratio of depth of surface segregation to nanoparticle size.

Analytical estimation of such surface effects is not
straightforward and will be done elsewhere\cite{Segregation}. Here
we only describe the source of the asymmetry observed in Fig.
\ref{Fig:PP_c}. Namely, it is observed that in the presence of a
free surface the Connolly-Williams bulk mixing potentials result
in weak segregation of Fe atoms to the surface. This causes a
depletion of Fe in the interior of the nanoparticle.  Thus the
effective concentration of Fe atoms within the nanoparticle but
outside the surface segregation region will be less than one half.
Therefore it is necessary to add Fe atoms to the interior of the
nanoparticle to achieve equiatomic composition outside the
segregating surface and accordingly to achieve the maximum L1$_0$
order in that region. That is why the maximum of the L1$_0$ order
parameter lies on the Fe-rich side of equiatomic composition in
Fig. \ref{Fig:PP_c}.  A similar effect (but for the pseudo phase
transition temperature in finite systems\cite{NoFinitePhTr}) was
observed in Ref. \onlinecite{Seagate} for the case in which a
strong segregation potential leading to the segregation of Pt to
the surface was introduced during Monte Carlo simulation. That
additional potential also results in a decrease of total order
parameter in accordance with our description above.

For nanoparticles chemically synthesized by methods such as the
"hot soap process"\cite{Sun}, the problem of the effect of the
surface on the interatomic exchange potentials is even more
complicated because these nanoparticles are likely to have unknown
atoms and molecules attached to their surfaces.  It should also be
noted that truncation of the mixing potentials at a surface is an
uncontrolled approximation.  The fact that our simulations yield
segregation of Fe to the surface should not be taken as evidence
that this would happen for a real nanoparticle.

\section{Conclusions and discussion}

In the present paper, the pair mixing potentials (which determine
the alloy configurational behavior) were calculated up to the third
coordination shell for equiatomic FePt alloys from first principles
using the Connolly-Williams method (Sec. \ref{CW}). It was
shown that the application of this method does not necessarily lead
to significant non-pair interactions if all of the structures
employed in the Connolly-Williams method correspond to the same
composition (canonical cluster expansion formalism\cite{CCE}).
The mixing potential was also calculated by adding the
strain-induced part of the interactions to mixing potentials
obtained from the KKR-CPA approach of Ref. \onlinecite{Staunton03}
(Sec. \ref{CPA}). The application of the Connolly-Williams method at
fixed composition allowed the direct comparison of these potentials
with those obtained from the KKR-CPA (plus strain-induced part). The
mixing potentials are shown to be similar with both giving values
for the order-disorder phase transition very close to the
experimental one. It was demonstrated that the Connolly-Williams
potentials (based on completely \emph{ordered} states) and the
KKR-CPA potentials (corresponding to completely \emph{disordered}
state) can be brought into very close correspondence to each
other simply by increasing the magnitude of the strain-induced
interactions added to KKR-CPA potential (Sec. \ref{CPA}).

It should be emphasized that it was not clear {\it a priori} that
either the Connolly-Williams method or the CPA approach could  be
trusted for a system as complicated as FePt. Magnetic alloys are
problematic for the Connolly-Williams method because the mixing
potentials are strongly dependent on the magnetic state of the
system.  Thus the chemical order and magnetic structure are
expected to be intimately related. On the other hand the CPA
approach is based on concentration fluctuations in a hypothetical
high temperature disordered state.  We consider the fact that both
methods are able to approximate the order-disorder temperature
well without adjustable parameters and that the mixing potentials
derived from the two approaches are rather similar to be a
remarkable result.

It was shown by first-principles calculations that parallel L1$_0$
antiphase boundaries are not energetically favorable (Sec.
\ref{APB}). This fact demonstrates the inapplicability of the
nearest neighbor interaction model in accordance with
Connolly-Williams and KKR-CPA results. The absence of parallel
L1$_0$ antiphase boundaries was confirmed by Monte Carlo simulation
using the Connolly-Williams potential (Secs. \ref{Bulk phase
transition}-\ref{nanoMC}).

In Secs. \ref{Bulk phase transition}-\ref{nanoMC}, the mixing
potentials obtained by the  Connolly-Williams method were
used to investigate the dependence of equilibrium L1$_0$ ordering on
temperature for bulk and for spherical nanoparticles with diameters
of 5.94nm, 3.5nm, 3.0nm and 2.5nm.  These calculations used both
Monte Carlo simulation and the analytical ring approximation. The
calculated order-disorder temperature for bulk (1495-1514 K) was in
relatively good agreement (4\% error) with the experimental value
(1572K). For nanoparticles of finite size, the (long range) order
parameter changed continuously from unity to finite asymptotic
values with increasing temperature. The nonzero asymptotic values
are a consequence of thermodynamic fluctuations of the order
parameter and are proportional to $1/\sqrt{N}$ where $N$ is the
number of atoms in a sample. Rather than a discontinuity indicative
of a phase transition we obtained an inflection point (a precursor
of a phase transition at large size) in the L1$_0$ order as a
function of temperature. This inflection point occurred at a
temperature below the bulk phase transition temperature and which
decreased as the particle size decreased.

According to our investigations, the experimental absence of
(relatively) high order in nanoparticles below 600$^{\circ}$C is
primarily a \emph{kinetic} problem rather than an equilibrium one.
For example, our calculations predict, that 3.5nm diameter particles
in configurational equilibrium at 600$^{\circ}$C (a typical
annealing temperature for promoting L1$_0$ ordering) have an L1$_0$
order parameter of 0.84 (compared to a maximum possible value equal
to unity). It should be noted that to rapidly obtain the correct
equilibrium state, we used simplified kinetics in our Monte Carlo
simulation. Namely, we allowed \emph{any} two randomly chosen atoms
to exchange their positions \emph{without} an additional diffusion
barrier. In a real alloy, the main mechanism of atomic diffusion is
much slower because it consists of the exchange of the positions of
atoms and neighboring vacancies through energy barriers. Moreover,
at each temperature we started the simulation from the completely
ordered state, whereas the actual nanoparticles are initially
prepared in disordered state and transformation from the disordered
to the ordered state may be much slower than the reverse one,
especially at low temperatures. Nevertheless, even with our
simplified kinetics, we observed a slowing down problem in
approaching the equilibrium ordered state at low temperatures. In
real nanoparticles this problem must be much worse. Kinetic
acceleration methods such as irradiation and/or addition of other
types of atoms may be useful in accelerating the formation of long
range order. Harrell et al. \cite{harrell} {\em have} observed
ordering on annealing at temperatures somewhat below 400$^\circ$C by
adding Ag or Au to the particles in the synthesis step.  However, it
appears that this ordering is also accompanied by agglomeration and
sintering of the particles. It should be emphasized that all of the
results presented in this paper correspond to equilibrium states and
are, therefore, independent of the particular kinetic pathways that
lead to these states.

In Sec. \ref{nanoMC}, it was demonstrated that for small finite
samples even for a composition independent, pair mixing potential
one obtains an asymmetric order parameter as a function of
composition with respect to equiatomic composition. In particular,
for FePt nanoparticles, if we neglect the effect of the surface on
the mixing potentials, the maximum of the L1$_0$ order parameter as
a function of composition is shifted from $c=0.5$ toward the Fe rich
region. The connection of such an asymmetry with surface segregation
is discussed in Sec. \ref{Surface effects}.

\section{Acknowledgments}

This research was supported by the Defense
Advanced Research Projects Agency through ONR contract
N00014-02-01-0590 and by the National Science Foundation through MRSEC Grant No.
DMR0213985 and Grant No. NSF SA4130-10092PG/82939. The authors thank Prof. J.W. Harrell  for stimulating
discussions and Dr. J. Velev for participation in early stages of
the present work. Special thanks to Prof. J. B. Staunton for
providing us the numerical values of KKR-CPA-based mixing potential
calculated in Ref. \onlinecite{Staunton03}. We are grateful to Prof.
M. Asta and Dr. O. N. Mryasov for communicating their results prior
to publication.

\appendix

\section{Definition of structures}
\label{SCW}

The present appendix (Tab. \ref{TAB:SCW} and Subsecs. 1-28)
defines the structures used in this paper. Unit cell basis vectors
are given in Cartesian coordinates in units of the fcc lattice
parameter. The Cartesian axes are directed along the edges of the
fcc cube. The third unit cell basis vector is perpendicular to the
first and second ones for all structures. All of the structures
have equiatomic composition $c=0.5$. The atom positions in the
tables are before relaxation. In addition to numbering, we also
include the structure designations that we used during our
calculations. For the first, second and twenty fourth structures
we put also the designations known in the literature ("L1$_0$",
"CH(40)" and "Z2", respectively).\cite{ZungerSCW} The first 23
structures are linearly independent and were used in the
Connolly-Williams method in Sec. \ref{CW}. The rest 5 structures
are linearly dependent (but not identical) to the first 23 ones.
They were considered in order to check the predictive power of the
Connolly-Williams potentials.

\begin{table}[htbp]
\caption{The characteristics of the structures used in the paper.
$i$ is the structure number, $N_{\texttt{u.c.}}^{\texttt{AB}}$ is
the total number of atoms in the unit cell of structure,
$E_{\texttt{u.c.}}$ is the energy of structure per its unit cell (as
obtained from VASP calculations), $V^{(0)}_{\texttt{u.c.}}$ is the
energy (per unit cell) of the structure with the same unit cell but
with Fe atoms only (see Sec. \ref{LG}), $S_{is}$ are the structural
coefficients ($i$=1,2...,28, $s$=1,2...,8) in the Connolly-Williams
system of equations (\ref{E3}). Note that the values of
$\varepsilon_i$ [see Eqs. (\ref{E3})-(\ref{Ei})] can be determined
from this table as
$\varepsilon_i=(E_{\texttt{u.c.}}-V^{(0)}_{\texttt{u.c.}})/N_{\texttt{u.c.}}^{\texttt{AB}}$,
where the values on the right side of this equation correspond to
the $i$-th structure.} \label{TAB:SCW}

\begin{tabular}{cccc|cccccccc}
$i$ & $N_{\texttt{u.c.}}^{\texttt{AB}}$ &  $V^{(0)}_{\texttt{u.c.}}$ & $E_{\texttt{u.c.}}$ &  %
                    \multicolumn{8}{c}{$S_{is} \times N_{\texttt{u.c.}}^{\texttt{AB}}$} \\ \hline
  1 & 16 & -129.0729 & -117.0299 &  16 &  24 &  32 &  48 &  32 &  32 &  64 &  24 \\
  2 & & & -116.1764 &  16 &  16 &  64 &  16 &  32 &   0 & 128 &  24 \\
  3 & & & -116.6366 &  16 &  20 &  48 &  32 &  32 &  16 &  96 &  16 \\
  4 & & & -114.5503 &  40 &  20 &  64 &  32 &  64 &  16 &  96 &  16 \\
  5 & & & -114.9625 & 34 & 20 & 60 & 32 & 54 & 16 & 92 & 18 \\
  6 & & & -114.6047 & 36 & 20 & 64 & 32 & 56 & 16 & 96 & 16 \\
  7 & 16 & -128.9796 &  -115.4659 & 27 & 14 & 38 & 24 & 54 &  8 & 76 & 24 \\
  8 & & & -115.3484 & 28 & 14 & 40 & 16 & 56 &  8 & 80 & 24 \\
  9 & 18 &  -145.2008 &  -130.8012 & 22 & 23 & 44 & 42 & 36 & 28 & 88 & 23 \\
 10 & & & -130.8094 & 22 & 23 & 40 & 42 & 44 & 28 & 84 & 23 \\
 11 & & & -131.0869 & 20 & 23 & 44 & 42 & 42 & 28 & 84 & 23 \\
 12 & 10 &   -80.6305 &   -72.2655 & 14 &  9 & 26 & 12 & 34 &  8 & 66 &  9 \\
 13 & 8 &   -64.5151 &   -57.1224 & 18 &  8 & 28 & 12 & 20 &  8 & 40 &  4 \\
 14 & & &  -57.7453 & 13 &  6 & 26 & 10 & 22 & 12 & 44 &  8 \\
 15 & 6 &   -48.3836 &   -43.0543 & 12 &  5 & 16 &  8 & 12 &  8 & 32 &  5 \\
 16 & 12 &   -96.7403 &   -85.1600 & 30 & 14 & 52 & 24 & 44 & 16 & 80 & 10 \\
 17  & & & -85.7956 & 25 & 12 & 46 & 20 & 38 & 16 & 72 & 10 \\
 18 & 12 &   -96.7380 &  -86.9505  & 16 & 10 & 32 & 16 & 40 & 16 & 80 & 10 \\
 19 & 12 &   -96.7780 &  -86.6918 & 20 & 10 & 30 & 18 & 38 &  8 & 58 & 14 \\
 20 & 8 &   -64.5043 &   -58.1203 & 10 &  8 & 20 & 12 & 28 &  8 & 56 &  4 \\
 21 & 20 &  -161.2637 &  -144.2564 & 35 & 18 & 47 & 24 & 74 &  8 & 93 & 26 \\
 22  & & & -144.1195 & 36 & 18 & 46 & 22 & 76 &  6 & 94 & 26 \\
 23 & 16 &  -129.0066 &  -115.4983 & 28 & 14 & 38 & 22 & 54 &  8 & 76 & 20 \\
 24 & 8 &   -64.5283 &   -57.6927 & 16 &  8 & 16 &  8 & 32 &  0 & 32 & 12 \\
 25 & & &  -57.8895 & 12 & 10 & 16 & 16 & 24 &  8 & 32 & 12 \\
 26 & & &  -57.7018 & 12 &  8 & 24 &  8 & 24 &  0 & 48 & 12 \\
 27 & & &  -58.1095 & 10 & 10 & 20 & 16 & 20 &  8 & 40 & 12 \\
 28 & & &  -58.2748 &  8 & 10 & 24 & 16 & 16 &  8 & 48 & 12
\end{tabular}
\end{table}

\subsection{4cub(1), L1$_0$}

Unit cell basis vectors Cartesian coordinates:

(1.0,0.0,0.0), (0.0,1.0,0.0), (0.0,0.0,4.0).

Atomic Cartesian coordinates

A:
(0.0,0.0,0.0),
(0.5,0.5,0.0),
(0.0,0.0,1.0),
(0.5,0.5,1.0),
(0.0,0.0,2.0),
(0.5,0.5,2.0),
(0.0,0.0,3.0),
(0.5,0.5,3.0);

B:
(0.5,0.0,0.5),
(0.0,0.5,0.5),
(0.5,0.0,1.5),
(0.0,0.5,1.5),
(0.5,0.0,2.5),
(0.0,0.5,2.5),
(0.5,0.0,3.5),
(0.0,0.5,3.5).

\subsection{4cub(2), APB(1), CH(40)}

Unit cell basis vectors Cartesian coordinates:

(1.0,0.0,0.0), (0.0,1.0,0.0), (0.0,0.0,4.0).

Atomic Cartesian coordinates

A: (0.0,0.0,0.0), (0.5,0.0,0.5), (0.5,0.5,1.0), (0.0,0.5,1.5),
(0.0,0.0,2.0), (0.5,0.0,2.5), (0.5,0.5,3.0), (0.0,0.5,3.5);

B: (0.5,0.5,0.0), (0.0,0.5,0.5), (0.0,0.0,1.0), (0.5,0.0,1.5),
(0.5,0.5,2.0), (0.0,0.5,2.5), (0.0,0.0,3.0), (0.5,0.0,3.5).

\subsection{4cub(3), APB(2)}

Unit cell basis vectors Cartesian coordinates:

(1.0,0.0,0.0), (0.0,1.0,0.0), (0.0,0.0,4.0).

Atomic Cartesian coordinates

A: (0.0,0.0,0.0), (0.5,0.0,0.5), (0.0,0.0,1.0), (0.5,0.0,1.5),
(0.5,0.5,2.0), (0.0,0.5,2.5), (0.5,0.5,3.0), (0.0,0.5,3.5);

B: (0.5,0.5,0.0), (0.0,0.5,0.5), (0.5,0.5,1.0), (0.0,0.5,1.5),
(0.0,0.0,2.0), (0.5,0.0,2.5), (0.0,0.0,3.0), (0.5,0.0,3.5).

\subsection{4cub(4)}

Unit cell basis vectors Cartesian coordinates:

(1.0,0.0,0.0), (0.0,1.0,0.0), (0.0,0.0,4.0).

Atomic Cartesian coordinates

A: (0.0,0.0,0.0), (0.5,0.5,0.0), (0.5,0.0,0.5), (0.0,0.5,0.5),
(0.0,0.0,1.0), (0.5,0.5,1.0), (0.5,0.0,1.5), (0.0,0.5,1.5);

B: (0.0,0.0,2.0), (0.5,0.5,2.0), (0.5,0.0,2.5), (0.0,0.5,2.5),
(0.0,0.0,3.0), (0.5,0.5,3.0), (0.5,0.0,3.5), (0.0,0.5,3.5).

\subsection{4cub(5)}

Unit cell basis vectors Cartesian coordinates:

(1.0,0.0,0.0), (0.0,1.0,0.0), (0.0,0.0,4.0).

Atomic Cartesian coordinates

A: (0.0,0.0,0.0), (0.5,0.5,0.0), (0.5,0.0,0.5), (0.0,0.5,0.5),
(0.0,0.0,1.0), (0.5,0.5,1.0), (0.5,0.0,1.5), (0.0,0.0,2.0);

B: (0.0,0.5,1.5), (0.5,0.5,2.0), (0.5,0.0,2.5), (0.0,0.5,2.5),
(0.0,0.0,3.0), (0.5,0.5,3.0), (0.5,0.0,3.5), (0.0,0.5,3.5).

\subsection{4cub(6)}

Unit cell basis vectors Cartesian coordinates:

(1.0,0.0,0.0), (0.0,1.0,0.0), (0.0,0.0,4.0).

Atomic Cartesian coordinates

A: (0.0,0.0,0.0), (0.5,0.5,0.0), (0.5,0.0,0.5), (0.0,0.5,0.5),
(0.0,0.0,1.0), (0.5,0.5,1.0), (0.5,0.0,1.5), (0.0,0.5,3.5);

B: (0.0,0.5,1.5), (0.0,0.0,2.0), (0.5,0.5,2.0), (0.5,0.0,2.5),
(0.0,0.5,2.5), (0.0,0.0,3.0), (0.5,0.5,3.0), (0.5,0.0,3.5).

\subsection{4ppd1-2pl(1)}

Unit cell basis vectors Cartesian coordinates:

(1.0,-1.0,0.0), (1.0,1.0,0.0), (0.0,0.0,2.0).

Atomic Cartesian coordinates

A: (0.0,0.0,0.0), (0.5,-0.5,0.0), (0.5,0.5,0.0), (1.0,0.0,0.0),
(0.5,0.0,0.5), (1.0,-0.5,0.5), (1.0,0.5,0.5), (0.0,0.0,1.0);

B: (1.5,0.0,0.5), (0.5,-0.5,1.0), (0.5,0.5,1.0), (1.0,0.0,1.0),
(0.5,0.0,1.5), (1.0,-0.5,1.5), (1.0,0.5,1.5), (1.5,0.0,1.5).

\subsection{4ppd1-2pl(2)}

Unit cell basis vectors Cartesian coordinates:

(1.0,-1.0,0.0), (1.0,1.0,0.0), (0.0,0.0,2.0).

Atomic Cartesian coordinates

A: (0.0,0.0,0.0), (0.5,-0.5,0.0), (0.5,0.5,0.0), (1.0,0.0,0.0),
(0.5,0.0,0.5), (1.0,-0.5,0.5), (1.0,0.5,0.5), (0.5,0.0,1.5);

B: (1.5,0.0,0.5), (0.0,0.0,1.0), (0.5,-0.5,1.0), (0.5,0.5,1.0),
(1.0,0.0,1.0), (1.0,-0.5,1.5), (1.0,0.5,1.5), (1.5,0.0,1.5).

\subsection{9ppd1-1pl(1)}

Unit cell basis vectors Cartesian coordinates:

(1.5,-1.5,0.0), (1.5,1.5,0.0), (0.0,0.0,1.0).

Atomic Cartesian coordinates

A: (0.0,0.0,0.0), (0.5,-0.5,0.0), (1.0,-1.0,0.0), (0.5,0.5,0.0),
(1.0,0.0,0.0), (1.5,-0.5,0.0), (1.0,1.0,0.0), (1.5,0.5,0.0),
(0.5,0.0,0.5);

B: (2.0,0.0,0.0), (1.0,-0.5,0.5), (1.5,-1.0,0.5), (1.0,0.5,0.5),
(1.5,0.0,0.5), (2.0,-0.5,0.5), (1.5,1.0,0.5), (2.0,0.5,0.5),
(2.5,0.0,0.5).

\subsection{9ppd1-1pl(2)}

Unit cell basis vectors Cartesian coordinates:

(1.5,-1.5,0.0), (1.5,1.5,0.0), (0.0,0.0,1.0).

Atomic Cartesian coordinates

A: (0.0,0.0,0.0), (0.5,-0.5,0.0), (1.0,-1.0,0.0), (0.5,0.5,0.0),
(1.0,0.0,0.0), (1.5,-0.5,0.0), (1.0,1.0,0.0), (1.5,0.5,0.0),
(1.0,-0.5,0.5);

B: (2.0,0.0,0.0), (0.5,0.0,0.5), (1.5,-1.0,0.5), (1.0,0.5,0.5),
(1.5,0.0,0.5), (2.0,-0.5,0.5), (1.5,1.0,0.5), (2.0,0.5,0.5),
(2.5,0.0,0.5).

\subsection{9ppd1-1pl(3)}

Unit cell basis vectors Cartesian coordinates:

(1.5,-1.5,0.0), (1.5,1.5,0.0), (0.0,0.0,1.0).

Atomic Cartesian coordinates

A: (0.0,0.0,0.0), (0.5,-0.5,0.0), (1.0,-1.0,0.0), (0.5,0.5,0.0),
(1.0,0.0,0.0), (1.5,-0.5,0.0), (1.0,1.0,0.0), (1.5,0.5,0.0),
(1.5,0.0,0.5);

B: (2.0,0.0,0.0), (0.5,0.0,0.5), (1.0,-0.5,0.5), (1.5,-1.0,0.5),
(1.0,0.5,0.5), (2.0,-0.5,0.5), (1.5,1.0,0.5), (2.0,0.5,0.5),
(2.5,0.0,0.5).

\subsection{1ppd2-1pl}

Unit cell basis vectors Cartesian coordinates:

(1.5,-0.5,0.0), (0.5,1.5,0.0), (0.0,0.0,1.0).

Atomic Cartesian coordinates

A: (0.0,0.0,0.0), (0.5,0.0,0.5), (0.5,0.5,0.0), (1.0,0.0,0.0),
(1.0,0.5,0.5);

B: (0.5,1.0,0.5), (1.0,1.0,0.0), (1.5,0.0,0.5), (1.5,0.5,0.0),
(1.5,1.0,0.5).

\subsection{2ppd3a-1pl(1)}

Unit cell basis vectors Cartesian coordinates:

(3.0,-1.0,0.0), (0.5,0.5,0.0), (0.0,0.0,1.0).

Atomic Cartesian coordinates

A: (0.0,0.0,0.0), (0.5,0.0,0.5), (1.0,0.0,0.0), (1.5,0.0,0.5);

B: (1.5,-0.5,0.0), (2.0,-0.5,0.5), (2.5,-0.5,0.0), (3.0,-0.5,0.5).

\subsection{2ppd3a-1pl(2)}

Unit cell basis vectors Cartesian coordinates:

(3.0,-1.0,0.0), (0.5,0.5,0.0), (0.0,0.0,1.0).

Atomic Cartesian coordinates

A: (0.0,0.0,0.0), (0.5,0.0,0.5), (1.0,0.0,0.0), (2.0,-0.5,0.5);

B: (1.5,0.0,0.5), (1.5,-0.5,0.0), (2.5,-0.5,0.0), (3.0,-0.5,0.5).

\subsection{1ppd4-1pl}

Unit cell basis vectors Cartesian coordinates:

(2.0,-1.0,0.0), (0.5,0.5,0.0), (0.0,0.0,1.0).

Atomic Cartesian coordinates

A: (0.0,0.0,0.0), (0.5,0.0,0.5), (1.0,0.0,0.0);

B: (1.0,-0.5,0.5), (1.5,-0.5,0.0), (2.0,-0.5,0.5).

\subsection{2ppd4a-1pl(1)}

Unit cell basis vectors Cartesian coordinates:

(4.0,-2.0,0.0), (0.5,0.5,0.0), (0.0,0.0,1.0).

Atomic Cartesian coordinates

A: (0.0,0.0,0.0), (0.5,0.0,0.5), (1.0,0.0,0.0), (1.0,-0.5,0.5),
(1.5,-0.5,0.0), (2.0,-0.5,0.5);

B: (2.0,-1.0,0.0), (2.5,-1.0,0.5), (3.0,-1.0,0.0), (3.0,-1.5,0.5),
(3.5,-1.5,0.0), (4.0,-1.5,0.5).

\subsection{2ppd4a-1pl(2)}

Unit cell basis vectors Cartesian coordinates:

(4.0,-2.0,0.0), (0.5,0.5,0.0), (0.0,0.0,1.0).

Atomic Cartesian coordinates

A: (0.0,0.0,0.0), (0.5,0.0,0.5), (1.0,0.0,0.0), (1.0,-0.5,0.5),
(1.5,-0.5,0.0), (2.5,-1.0,0.5);

B: (2.0,-0.5,0.5), (2.0,-1.0,0.0), (3.0,-1.0,0.0), (3.0,-1.5,0.5),
(3.5,-1.5,0.0), (4.0,-1.5,0.5).

\subsection{2ppd4b-1pl}

Unit cell basis vectors Cartesian coordinates:

(2.0,-1.0,0.0), (1.0,1.0,0.0), (0.0,0.0,1.0).

Atomic Cartesian coordinates

A: (0.0,0.0,0.0), (0.5,0.0,0.5), (1.0,0.0,0.0), (1.0,-0.5,0.5),
(1.5,-0.5,0.0), (2.0,-0.5,0.5);

B: (0.5,0.5,0.0), (1.0,0.5,0.5), (1.5,0.5,0.0), (1.5,0.0,0.5),
(2.0,0.0,0.0), (2.5,0.0,0.5).

\subsection{1ppd4-2pl}

Unit cell basis vectors Cartesian coordinates:

(2.0,-1.0,0.0), (0.5,0.5,0.0), (0.0,0.0,2.0).

Atomic Cartesian coordinates

A: (0.0,0.0,0.0), (0.5,0.0,0.5), (1.0,0.0,0.0), (1.0,-0.5,0.5),
(1.5,-0.5,0.0), (0.0,0.0,1.0);

B: (2.0,-0.5,0.5), (0.5,0.0,1.5), (1.0,0.0,1.0), (1.0,-0.5,1.5),
(1.5,-0.5,1.0), (2.0,-0.5,1.5).

\subsection{2ppd3b-1pl}

Unit cell basis vectors Cartesian coordinates:

(1.5,-0.5,0.0), (1.0,1.0,0.0), (0.0,0.0,1.0).

Atomic Cartesian coordinates

A: (0.0,0.0,0.0), (0.5,0.0,0.5), (1.0,0.0,0.0), (1.5,0.0,0.5);

B: (0.5,0.5,0.0), (1.0,0.5,0.5), (1.5,0.5,0.0), (2.0,0.5,0.5).

\subsection{1ppd2-2pl(1)}

Unit cell basis vectors Cartesian coordinates:

(1.5,-0.5,0.0), (0.5,1.5,0.0), (0.0,0.0,2.0).

Atomic Cartesian coordinates

A: (0.0,0.0,0.0), (0.5,0.0,0.5), (0.5,0.5,0.0), (0.5,1.0,0.5),
(1.0,0.0,0.0), (1.0,0.5,0.5), (1.0,1.0,0.0), (1.5,0.0,0.5),
(1.5,0.5,0.0), (0.0,0.0,1.0);

B: (1.5,1.0,0.5), (0.5,0.0,1.5), (0.5,0.5,1.0), (0.5,1.0,1.5),
(1.0,0.0,1.0), (1.0,0.5,1.5), (1.0,1.0,1.0), (1.5,0.0,1.5),
(1.5,0.5,1.0), (1.5,1.0,1.5).

\subsection{1ppd2-2pl(2)}

Unit cell basis vectors Cartesian coordinates:

(1.5,-0.5,0.0), (0.5,1.5,0.0), (0.0,0.0,2.0).

Atomic Cartesian coordinates

A: (0.0,0.0,0.0), (0.5,0.0,0.5), (0.5,0.5,0.0), (0.5,1.0,0.5),
(1.0,0.0,0.0), (1.0,0.5,0.5), (1.0,1.0,0.0), (1.5,0.0,0.5),
(1.5,0.5,0.0), (0.5,0.0,1.5);

B: (1.5,1.0,0.5), (0.0,0.0,1.0), (0.5,0.5,1.0), (0.5,1.0,1.5),
(1.0,0.0,1.0), (1.0,0.5,1.5), (1.0,1.0,1.0), (1.5,0.0,1.5),
(1.5,0.5,1.0), (1.5,1.0,1.5).

\subsection{2ppd3-3c-2pl}

Unit cell basis vectors Cartesian coordinates:

(3.5,-0.5,0.0), (0.5,0.5,0.0), (0.0,0.0,2.0).

Atomic Cartesian coordinates

A: (0.0,0.0,0.0), (0.5,0.0,0.5), (1.0,0.0,0.0), (1.5,0.0,0.5),
(2.0,0.0,0.0), (2.5,0.0,0.5), (3.0,0.0,0.0), (0.0,0.0,1.0);

B: (3.5,0.0,0.5), (0.5,0.0,1.5), (1.0,0.0,1.0), (1.5,0.0,1.5),
(2.0,0.0,1.0), (2.5,0.0,1.5), (3.0,0.0,1.0), (3.5,0.0,1.5).

\subsection{2cub(1), Z2}

Unit cell basis vectors Cartesian coordinates:

(1.0,0.0,0.0), (0.0,1.0,0.0), (0.0,0.0,2.0).

Atomic Cartesian coordinates

A:
(0.0,0.0,0.0),
(0.5,0.5,0.0),
(0.5,0.0,0.5),
(0.0,0.5,0.5);

B:
(0.0,0.0,1.0),
(0.5,0.5,1.0),
(0.5,0.0,1.5),
(0.0,0.5,1.5).

\subsection{2cub(2)}

Unit cell basis vectors Cartesian coordinates:

(1.0,0.0,0.0), (0.0,1.0,0.0), (0.0,0.0,2.0).

Atomic Cartesian coordinates

A:
(0.0,0.0,0.0),
(0.5,0.5,0.0),
(0.5,0.0,0.5),
(0.5,0.0,1.5);

B:
(0.0,0.5,0.5),
(0.0,0.0,1.0),
(0.5,0.5,1.0),
(0.0,0.5,1.5).

\subsection{2cub(3)}

Unit cell basis vectors Cartesian coordinates:

(1.0,0.0,0.0), (0.0,1.0,0.0), (0.0,0.0,2.0).

Atomic Cartesian coordinates

A:
(0.0,0.0,0.0),
(0.5,0.5,0.0),
(0.5,0.0,0.5),
(0.0,0.5,1.5);

B:
(0.0,0.5,0.5),
(0.0,0.0,1.0),
(0.5,0.5,1.0),
(0.5,0.0,1.5).

\subsection{2cub(4)}

Unit cell basis vectors Cartesian coordinates:

(1.0,0.0,0.0), (0.0,1.0,0.0), (0.0,0.0,2.0).

Atomic Cartesian coordinates

A:
(0.0,0.0,0.0),
(0.5,0.5,0.0),
(0.5,0.0,0.5),
(0.0,0.0,1.0);

B:
(0.0,0.5,0.5),
(0.5,0.5,1.0),
(0.5,0.0,1.5),
(0.0,0.5,1.5).

\subsection{2cub(5)}

Unit cell basis vectors Cartesian coordinates:

(1.0,0.0,0.0), (0.0,1.0,0.0), (0.0,0.0,2.0).

Atomic Cartesian coordinates

A:
(0.0,0.0,0.0),
(0.5,0.0,0.5),
(0.0,0.0,1.0),
(0.0,0.5,1.5);

B:
(0.5,0.5,0.0),
(0.0,0.5,0.5),
(0.5,0.5,1.0),
(0.5,0.0,1.5).


\begin{thebibliography}{*}

\bibitem[*]{Chep}Electronic address: r\_chepulskii@yahoo.com

\bibitem{Sun} S. Sun, C.B. Murray, D. Weller, L. Folks, and A. Moser, Science \textbf{287}, 1989
(2000).

\bibitem{Takahashi03} Y.K. Takahashi, T. Ohkubo, M. Ohnuma, and K. Hono, J. Appl. Phys. \textbf{93}, 7166
(2003).

\bibitem{Sint} Annealing at $\sim$600$^{\circ}$C results in sintering of
nanoparticles into larger agglomerates which is not desirable for
high density magnetic recording.

\bibitem{L10} L1$_0$ is an f.c.c. tetragonal structure, in which
atoms of two types form layers occupying alternating (001) or (010)
or (001) planes of the original f.c.c. lattice.

\bibitem{NoFinitePhTr} O.G. Mouritsen, \emph{Computer
Studies of Phase Transitions and Critical Phenomena}
(Springer-Verlag, Berlin, 1984), Sec. 2.2.8.

\bibitem{Short} Some preliminary results of the present paper were presented at
the Magnetism and Magnetic Materials Conference (2004) and will be
published in the Proceedings of the Conference: R.V. Chepulskii, J.
Velev, and W.H. Butler, J. Appl. Phys. \textbf{97} (2005) (pages
will be known soon).

\bibitem{CW}  J. W. D. Connolly and A. R. Williams, Phys. Rev. B \textbf {27},
5169 (1983);
D. de Fontaine, Solid State Physics \textbf{47}, 33 (1994);
A. Zunger, in \emph{Statics and Dynamics of Alloy
Phase Transformations}, edited by P. Turchi and A. Gonis (Plenum,
New York, 1994), pp. 361-419; A. Zunger, L. Wang, G. Hart, and M.
Sanati, Modell. Simul. Mater. Sci. Eng. \textbf{10}, 1 (2002);
S. M\"{u}ller, J. Phys.: Condens. Matter \textbf{15}, R1429 (2003);
V. Blum and A. Zunger, Phys. Rev. B \textbf{70}, 155108 (2004).

\bibitem{Ceder02} A. van de Walle and G. Ceder, Journal of Phase Equilibria \textbf{23}, 348 (2002).

\bibitem{CW-name} We use the term "Connolly-Williams method" to describe a procedure
for extracting mixing potential values from the known energies of a set of structures
acknowledging the authors who first applied this method.  Significant modifications 
have been made to this approach by later authors including the modifications described
in this paper.  Other names for this approach for obtaining mixing potentials 
are "structure inversion method" and "cluster expansion method"\cite{CW,Ceder02}. 

\bibitem{Ducast} F. Ducastelle, \emph{Order and Phase
Stability in Alloys} (Elsevier, New York, 1991).

\bibitem{VASP} G. Kresse and J. Furthmüller, Comput. Mater. Sci. \textbf{6}, 15
(1996); G. Kresse and J. Furthmüller, Phys. Rev. B \textbf{54}, 11169 (1996);
G. Kresse and D. Joubert, Phys. Rev. B \textbf{59}, 1758 (1999).

\bibitem{Gyorffy-83}  B. L. Gy\"{o}rffy and G. M. Stocks, Phys. Rev. Lett.
{\textbf 50}, 374 (1983).

\bibitem{Johnson-90}  D. D. Johnson, D. M. Nicholson, F. J. Pinski, B. L.
Gy\"{o}rffy and G. M. Stocks, Phys. Rev. B {\textbf 41}, 9701,
(1990).

\bibitem{Staunton-94}  J. B. Staunton, D. D. Johnson, and F. J. Pinski,
Phys. Rev. B {\textbf 50}, 1450 (1994).

\bibitem{Pinski-98}  F. J. Pinski, J. B. Staunton and D. D. Johnson, Phys.
Rev. B {\textbf 57}, 15177 (1998).

\bibitem{Staunton03} S. Ostanin, S. S. A. Razee, J. B. Staunton, B. Ginatempo, and Ezio
Bruno, J. Appl. Phys. \textbf{93}, 453 (2003).

\bibitem{Metropolis} N. Metropolis, A.W. Rosenbluth, M.N. Rosenbluth, A.H.
Teller, and E. Teller, J. Chem. Phys. 21, 1087 (1953); M.E.J.
Newman and G.T. Barkema, \emph{Monte Carlo Methods in Statistical
Physics} (Oxford University Press, Oxford, 1999), Sec.
3.1.

\bibitem{Ring} R.V. Chepulskii, Solid State Commun. \textbf{115}, 497
(2000); Phys. Rev. B \textbf{69}, 134431 (2004).

\bibitem{Lee52}T. D. Lee and C. N. Yang, Phys. Rev. \textbf
{87}, 410 (1952).

\bibitem{Books}M. A. Krivoglaz and A. A. Smirnov, \emph{The Theory of Order-Disorder
in Alloys} (Macdonald, London, 1964); A.G. Khachaturyan, Prog.
Mater. Sci. \textbf{22}, 1 (1978); D. de Fontaine, Solid State
Phys. \textbf{34}, 73 (1979).

\bibitem{Sym-I}  V. N. Bugaev and R. V. Chepulskii, Acta Cryst. A \textbf{51},
456 (1995).

\bibitem{CE} J. M. Sanchez, F. Ducastelle, and D. Gratias, Physica A \textbf{128}, 334 (1984).

\bibitem{CPA-V(c)} R. V. Chepulskii, J. B. Staunton, Ezio Bruno, B. Ginatempo, and D. D. Johnson,
Phys. Rev. B 65, 064201 (2001).

\bibitem{CCE} M. Asta, C. Wolverton, D. de Fontaine, and H. Dreyss\'{e}, Phys. Rev. B \textbf{44}, 4907 (1991);
C. Wolverton, M. Asta, H. Dreyss\'{e}, and D. de Fontaine, Phys.
Rev. B \textbf{44}, 4914 (1991).

\bibitem{Zark} N.A. Zarkevich, D.D. Johnson, and A.V. Smirnov, Acta Mater. \textbf{50}, 2443 (2002).

\bibitem{PhDiAsym}R. V. Chepulskii, J. Phys.: Condens. Matter \textbf{10},
1505 (1998), Sec. 5.

\bibitem{acExper} JCPDS-International Centre for Diffraction Data, 1999.

\bibitem{ZarkPRL} N. A. Zarkevich and D. D. Johnson, Phys. Rev. Lett. \textbf{92}, 255702 (2004).

\bibitem{CVmin} The use of that number of fitting parameters within the Connolly-Williams method
that provides a minimum of cross-validation error allows one to avoid ``under'' or ``over'' fitting and results in the highest predictive power of the Connolly-Williams method 
(see Sec. 3.1.1 in Ref. \onlinecite{Ceder02}).

\bibitem{higher_shells} We found
the mixing potential values for shells greater than the third one
to be so small that their accurate calculation requires an
increase in the accuracy of the VASP calculations and the
consideration of many more structures. It is important that the
accuracy of the Connolly-Williams structure energy fitting was
hardly affected by such an increase of the number of fitting
parameters (see Fig. \ref{Fig:CV}).

\bibitem{Noise} It should be noted that we found the bulk order-disorder phase transition temperature
to be sensitive to small changes in the energies of structures calculated from first principles.
For example, the change in $\varepsilon_i$ of 10 meV can cause phase transition temperature
to change by 50 K. Taking into account the approximations used at first principles calculations
within the VASP code (generalized gradient approximation to density-functional theory,
neglect of lattice vibrations and electronic excitations, etc.), one should realize
the corresponding error of statistical calculations based on those calculations.

\bibitem{Khacha} A. G. Khachaturyan, Fiz. Tverd. Tela \textbf{9}, 2861 (1967)
[Soviet Phys. Solid St. \textbf{9}, 2249 (1968)]; Prog. Mater.
Sci. \textbf{22}, 1 (1978); \emph{Theory of Structural
Transformations in Solids} (Wiley, New York, 1983).

\bibitem{Krivoglaz} M. A. Krivoglaz,  Zh. Eksp. Teor. Fiz. \textbf{34}, 204 (1958);
\emph{The Theory of X-Ray and Thermal Neutron Scattering From Real
Crystals} (Plenum, New York, 1969); \emph{Diffuse Scattering of
X-rays and Thermal Neutrons by Fluctuational Inhomogeneities of
Imperfect Crystals} (Springer, Berlin, 1996).

\bibitem{Mohri} Y. Chen, S. Iwata, and T. Mohri, Calphad \textbf{26}, 583
(2002).

\bibitem{Seagate} B. Yang, M. Asta, O. N. Mryasov, T. Klemmer, and R. Chantrell,
Scripta Materialia, in press.

\bibitem{Matsubara} T. J. Matsubara, J. Phys. Soc. Japan \textbf{7}, 270 (1952).

\bibitem{Kanzaki} H. Kanzaki, J. Phys. Chem. Solids \textbf{2}, 24 (1957).

\bibitem{Kushwaha} M. S. Kushwaha and S. S. Kushwaha, Phys. Status Sol. (b) \textbf{87}, 247 (1978).

\bibitem{Omegi} D. H. Dutton, B. N. Brockhouse, and A. P. Miiller,
Can. J. Phys. \textbf{50}, 2915 (1972).

\bibitem{ElastConsts} G. Simmons and H. Wang, \emph{Single Crystal Elastic
Constants ans Calculated Aggregate Properties: A Handbook}, 2nd
ed. (MIT Press, Cambridge, MA, 1971).

\bibitem{Pearson} W. B. Pearson, \emph{A Handbook of lattice spacing and structures in metals and alloys}
(Pergamon Press, Oxford, 1967) v1,2.

\bibitem{Bradley and Cracknell 1972} C. J. Bradley  and A. P. Cracknell,
\emph{The Mathematical Theory of Symmetry in Solids}
(University Press, Oxford, 1972).

\bibitem{V_PP} D. D. Johnson, A. V. Smirnov, J. B. Staunton, F. J. Pinski, and W. A. Shelton,
Phys. Rev. B \textbf{62}, R11917 (2000).

\bibitem{magneto} The additional error may be attributed to the different magnetic states
in completely oredered (ferromagnetic) and disordered (paramagnetic) states.

\bibitem{TB} Durga Paudyal, Tanusri Saha-Dasgupta and Abhijit Mookerjee, J. Phys.: Condens. Matter \textbf{16},  7247 (2004).

\bibitem{FePt-L10-bulk} \emph{Binary Alloy Phase Diagrams}, 2nd ed., edited by T. B. Massalski,
H. Okamoto, P. K. Subramanion, and L. Kacprzak (American
Society for Metals, Metals Park, OH, 1990).

\bibitem{Hansen} M. Hansen, \emph{Constitution of binary alloys} (New York, McGraw Hill, 1958).

\bibitem{sphere} Note that, in fact, the "spherical" nanoparticles used in the Monte Carlo simulation have
a trancated-octahedron geometry (see Ref. \onlinecite{Seagate}).

\bibitem{Cu3Au} T. Tadaki, T. Kinoshita, Y. Nakata, T. Ohkubo, and Y. Hirotsu, Z. Phys. D \textbf{40}, 493 (1997).

\bibitem{LandLifsh}L. D. Landau and E. M. Lifshitz, \emph{Course of Theoretical Physics}
(Pergamon, Oxford, 1980), Vol. 5.

\bibitem{Segregation} R.V. Chepulskii and W.H. Butler, in preparation.

\bibitem{harrell} S. Kang, D.E. Nikles and J. W. Harrell,  Nano Letters, \textbf{2}, 1033 (2002).

\bibitem{ZungerSCW} Z. W. Lu, S.–H. Wei,  A. Zunger, S. Frota–Pessoa, and L. G. Ferreira,
Phys. Rev. B \textbf{44}, 512 (1991).

\end{thebibliography}
\end{document}